\documentclass[ 
 amsmath,amssymb,
 aps,
prb,twocolumn,
longbibliography,
]{revtex4-2}

\usepackage{graphicx,bm,braket,color,hyperref,comment}
\usepackage{fancybox}
\usepackage{ulem}
\hypersetup{colorlinks=true,citecolor=blue,linkcolor=blue,urlcolor=blue}
\usepackage{dcolumn}
\usepackage{bbm}

\usepackage{amsfonts}
\usepackage{relsize}
\usepackage{here}

\newcommand{\bq}{ \bm{q} }
\newcommand{\QQ}{ {\boldsymbol \phi }}
\newcommand{\bk}{ {\bm{k}} }

\newcommand{\bH}{ \bm{H} }

\newcommand{\br}{ \bm{r} }
\newcommand{\bQ}{ \bm{Q} }

\newcommand{\nn}{ \nonumber }

\begin{document}


\title{L-point quadrupole order under magnetic field in cubic PrIr$_2$Zn$_{20}$
}

\author{Hitoko Okanoya}

\author{Kazumasa Hattori}%
 \email{hattori@tmu.ac.jp}
\affiliation{%
 Department of Physics, Tokyo Metropolitan University,\\ 1-1, Minami-osawa, Hachioji, Tokyo 192-0397, Japan
}%

\date{\today}

\begin{abstract}
We study quadrupole orders in heavy-fermion compound 
PrIr$_2$Zn$_{20}$ under magnetic fields on the basis of the Landau theory. 
Assuming $E_g$ electric quadrupolar orders in the cubic symmetry with the ordering wavenumber at the L points in the face-centered cubic lattice Brillouin zone as observed experimentally, we construct a Landau free energy and analyze the resulting free energy. We find that the unidentified high-temperature ordered phase under the magnetic field $\bH\parallel [001]$ reported earlier arises from the rotation of the quadrupole moments of f electrons on the Pr site. We also discuss the phase diagram for other magnetic-field directions and possible double-$\bq$ quadrupolar orders in this system. 
\end{abstract}

\maketitle


\section{Introduction}

Strongly correlated electron systems with orbital moments have attracted significant attention in recent years \cite{Kugel1973, Tokura2000}.
Various multipolar orders appear in d- and f-electron-based materials and have been intensively studied \cite{mydosh2014hidden,Kuramoto2009-ll,Chen2010}. Among them, electric quadrupole orders are the most typical and extensively investigated ones \cite{Shiina1997-ia,Onimaru2005-db,Chubukov2016-oy}. In addition, spin nematic orders have also been studied \cite{Andreev1984-vf,Lauchli2006-vf,Tsunetsugu2006-mf,Jiang2023-wi}. The nature of orbital degrees of freedom leads to spatially anisotropic interactions, and thus, the symmetry in orbital space is generally anisotropic. Each type of orbital order has unique properties, and understanding these properties remains a key issue in condensed matter theory and experiment.

Pr-based compounds Pr$T_2X_{20}$ ($T$ = Ir, Rh, Ti, V; $X$ = Zn, Al) form one of the most well-studied series of materials, commonly referred to as the ``1-2-20'' compounds \cite{OnimaruKusunose2016,Pottgen2023-bh}. The Pr ions form a diamond structure, and each Pr ion has a 4f$^2$ electron configuration with the cubic T$_d$ point group. Its ground state in the crystalline electric field (CEF) is a non-Kramers doublet that couples with conduction electrons. Since the excited CEF states are well separated from the ground-state doublet by more than 100 K \cite{Iwasa2013-dy}, these compounds provide an ideal setting for two-channel Kondo effects \cite{Cox1987-kc} to occur, leading to expectations of exotic properties arising from such effects \cite{Kusunose2016-sd,Inui2020-lu,Lenk2024-rr}. Indeed, while several compounds exhibit two-channel Kondo behavior \cite{Sakai2011}, others undergo orbital ordering at low temperatures. The phase transition can be modeled by interacting non-Kramers doublets, which support two quadrupole $\sim $\{$2z^2$$-$$x^2$$-$$y^2$, $\sqrt{3}($$x^2$$-$$y^2)$\} and one octupole ($T_{xyz}$) operators. Several theoretical studies have analyzed multipolar ordering in these systems. 
 Analyses by the mean-field theory \cite{Hattori2014,Ishitobi2019-du} for quadrupole moments and classical Monte Carlo (MC) simulations \cite{Hattori2016} were carried out in the early stage. Even for ferroic quadrupole orders in PrTi$_2$Al$_{20}$, the temperature-magnetic field phase diagrams are nontrivial and have been intensively studied \cite{Taniguchi2019-uo,Kittaka2020,Freyer2020-ns}. The realization of octupole order in PrV$_2$Al$_{20}$ \cite{Sakai2011,Ye2024-vd} has also been intensively studied, including approaches such as Landau theory \cite{Lee2018}, MC analysis \cite{Freyer2018}, and studies of multiple-$\bq$ multipolar ordering \cite{Ishitobi2021}.

Among the 1-2-20 compounds, PrIr$_2$Zn$_{20}$ is a prototypical material, exhibiting quadrupole ordering below 1 K and undergoing a phase transition to a superconducting state at 0.1 K \cite{Onimaru2011-of,Ishii2011-mj,onimaru2016}. The early neutron scattering experiment \cite{Iwasa2017-ni} under an applied magnetic field $\bm{H} \parallel [\bar{1}10]$ suggested an ordering wave number of $\bq=(\pi,\pi,\pi)$, corresponding to the L point at the first Brillouin zone boundary. Here, we have set the lattice constant to unity. The analysis of induced magnetic moments indicated  a  quadrupole order of the $x^2$$-$$y^2$ type for $\bm{H}\parallel [\bar{1}10]$. Recent reinvestigations of the temperature–magnetic field phase diagram have revealed strong anisotropy and highlight the existence of the high-temperature phase \cite{Kittaka2024-cn} reported in earlier studies \cite{Onimaru2011-of} across a wide range of magnetic field directions from [001] to [111], while it is absent for $\bm{H}\parallel [110]$.

In this paper, we analyze a phenomenological Landau model for PrIr$_2$Zn$_{20}$, incorporating $E_g$ quadrupole moments at the L points. We elucidate the nature of $E_g$ quadrupole ordering under magnetic fields and its dependence on field directions. We examine the order parameter suggested by the neutron scattering experiments for $\bm{H}\parallel [\bar{1}10]$, construct phase diagrams under applied magnetic fields, and compare them with the experimental results. The quadrupole orders at the L points reveal unique multiple-$\bq$ orders, which differ from those in the previous studies on multiple-$\bq$ quadrupole orders at the X-point ordering wavenumbers in the face-center-cubic (fcc) lattice \cite{Tsunetsugu2021,Hattori2023-pj}.

This paper is organized as follows. In Sec.~\ref{sec:model}, we construct the Landau free energy for the $E_g$ quadrupole moments at wavenumbers $\bq$$=$$(\pi,\pi,\pi)$ and $(0, 0, 0)$. In Sec.~\ref{sec:results}, we present the results of the Landau analysis for single-$\bq$ quadrupole orders, outlining the fundamental properties of the system. We then discuss multiple-$\bq$ quadrupole orders and the stability of various ordered phases in Sec.~\ref{sec:MultipleQ}. In Sec.~\ref{sec:dis}, we compare our results with experimental data for PrIr$_2$Zn$_{20}$ and related materials. Finally, Sec.~\ref{sec:sum} provides a summary of this paper.

\section{Model}\label{sec:model}
We begin by introducing the exchange interactions between the $E_g$ quadrupole moments in Sec.~\ref{ex} and the Landau free energy for the order parameters at the L points in Sec.~\ref{landau}. Mode-coupling terms related to the order parameter at the L points and other degrees of freedom are discussed in Sec.~\ref{modecoupling}.  Throughout this paper, we do not consider the $T_{xyz}$ octupolar degrees of freedom, since any octupolar orders have not been experimentally reported in PrIr$_2$Zn$_{20}$ so far. Although the excited CEF states possess various multipole moments such as magnetic dipole, octupole, and $T_{2}$-type quadrupole moments, we will ignore them since they are expected to play a minor role in PrIr$_2$Zn$_{20}$.


\begin{figure}[t!]
\begin{center}
\includegraphics[width=0.5\textwidth]{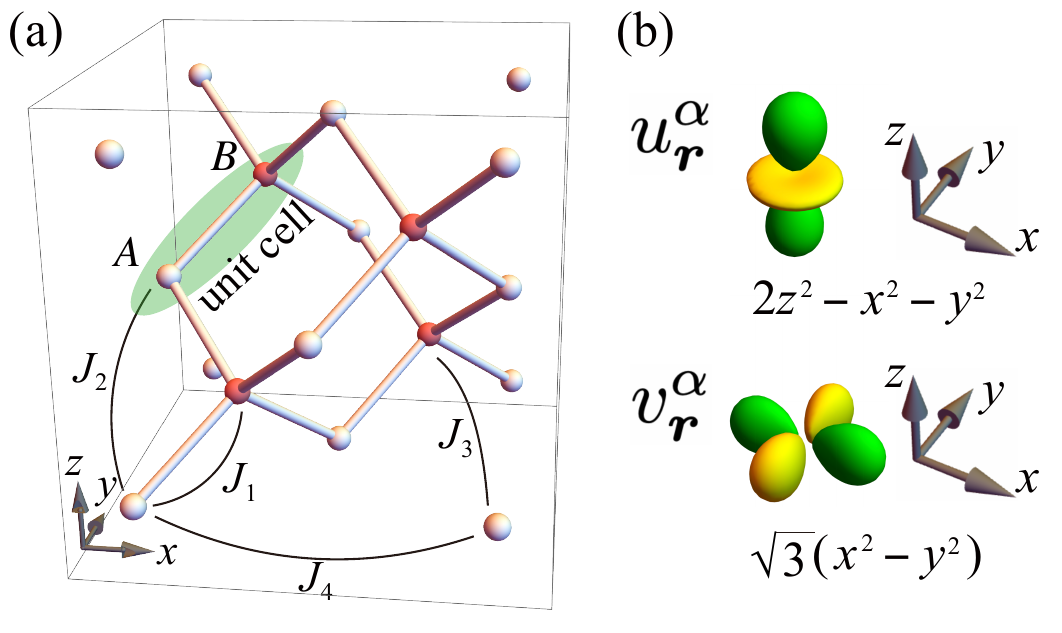}
\end{center}
\caption{(a) Crystal structure of the diamond structure formed by Pr ions in PrIr$_2$Zn$_{20}$. $A$ ($B$) sublattice sites are illustrated in white (red) spheres. The unit cell is taken as two sites along the [111] direction as indicated by shaded oval in green. Exchange interactions $J_1\sim J_4$ are also indicated. (b) Quadrupole moments $u_{\bm r}^\alpha$ and $v_{\bm r}^\alpha$ are schematically shown, where the different color represent the sign of moments.
}
\label{fig:0}
\end{figure}



\begin{figure}[t!]
\begin{center}
\includegraphics[width=0.45\textwidth]{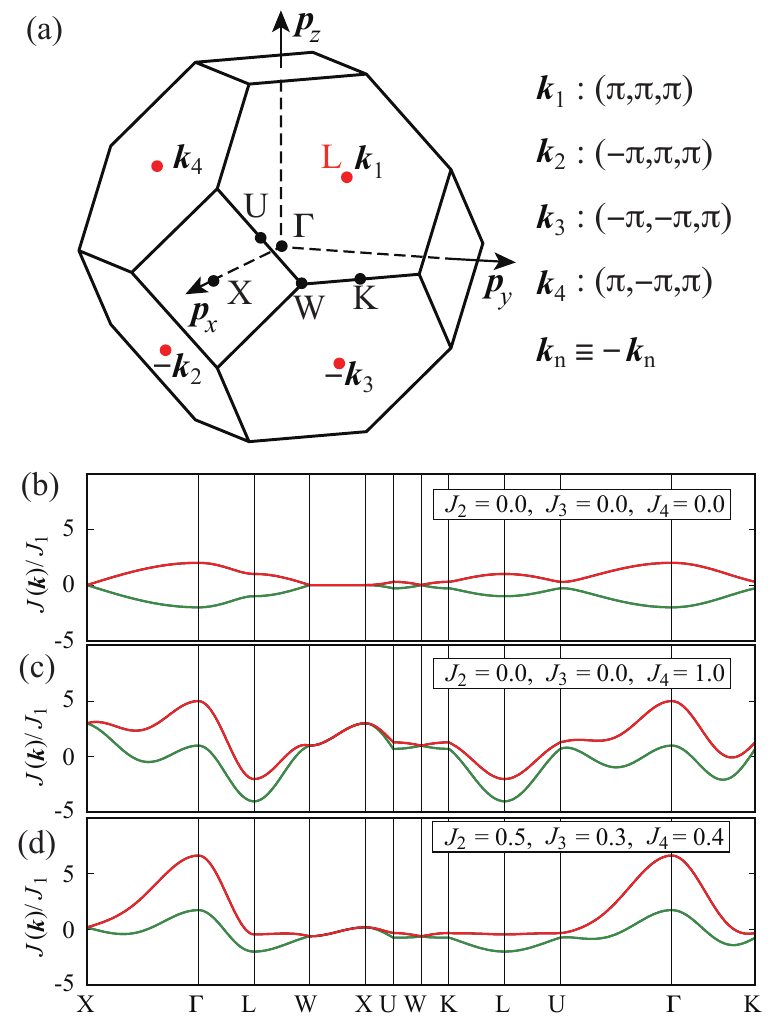}
\end{center}
\caption{(a) The first Brillouin zone of the fcc lattice. The four L points are indicated by filled red circles and other symmetric points are indicated by filled black circles.  (b)--(d) Fourier transform of the exchange interactions with $J_1=1$. (b) $(J_2,J_3,J_4)=(0.0,0.0,0.0)$, 
(c) $(0.0,0.0,1.0)$, and (d) $(0.5,0.3,0.4)$. }
\label{fig:1}
\end{figure}


\subsection{Exchange interaction}\label{ex}
Local moments of f electrons on the Pr sites, which form the diamond structure as shown in Fig.~\ref{fig:0}(a), tend to order at low temperatures. This is usually caused by Ruderman-Kittel-Kasuya-Yosida interaction due to the presence of the itinerant electrons in heavy-fermion compounds in general.
To discuss their ordering, it is reasonable to integrate the itinerant electrons and obtain the effective exchange model for the localized f electron moments. Since the local moments play a dominant role in the symmetry breaking orders in these compounds, we use the localized model as a minimal one to understand the nature of their symmetry breakings. 
Although it is desirable to obtain the information of the exchange interactions from the excitation spectra by inelastic x-ray or neutron scattering experiments, there is no data available  at present. Thus, we use a phenomenological and minimal model which realizes the L point order in this paper.  

Let us denote 
$E_g$ quadrupole moments at the position $\br=(x,y,z)$ in the diamond structure as $\bm{\phi}_{\br}^\alpha=(u_{\br}^{\alpha},v_{\br}^\alpha)$, where $\alpha=A$ or $B$ is the sublattice index. See Fig.~\ref{fig:0}(b). We take one of the $A$-sublattice site at the origin $\br=(x,y,z)=(0,0,0)$ and its pair $B$ sublattice site in the unit cell at $(\tfrac{1}{4},\tfrac{1}{4},\tfrac{1}{4})$, where we set the lattice constant $a=1$. $\bm{\phi}_{\br}^\alpha$ forms a two-dimensional vector field as a classical variable. The two components are  $u$ $\sim$ $2z^2$$-$$x^2$$-$$y^2$ and $v$ $\sim$ $\sqrt{3}$($x^2$$-$$y^2$). The Fourier transform in the wavenumber $\bk$ space is defined as
\begin{align}
{\boldsymbol \phi}_{\bk}^\alpha= 
	\begin{bmatrix}
		u^\alpha_{\bk}\\
		v^\alpha_{\bk}
	\end{bmatrix}=\frac{1}{\sqrt{N}}
	\sum_{\br} e^{-i\bk \cdot \br}
	\begin{bmatrix}
		u^\alpha_{\br}\\
		v^\alpha_{\br}
	\end{bmatrix}.	
\end{align} 
Here, the summation for ${\br}$ runs over the $A$-sublattice sites and $N$ is the number of the unit cell.

The exchange Hamiltonian is in general written as 
\begin{align}
	{\mathcal H}=\sum_{\bk}[u^A_{-\bk}, v^A_{-\bk}, u^B_{-\bk},v^B_{-\bk} ]
	\begin{bmatrix}
		{\mathcal H}^{AA}_{\bk} & {\mathcal H}^{AB}_{\bk}\\
		{{\mathcal H}^{AB}_{\bk}}^\dag & {\mathcal H}^{AA}_{\bk}
	\end{bmatrix}
	\begin{bmatrix}
		u^A_{\bk}\\
		v^A_{\bk}\\
		u^B_{\bk}\\
		v^B_{\bk}
	\end{bmatrix}\label{eq:H}
	,
\end{align}
where ${\mathcal H}^{\alpha\beta}_{\bk}$ is a $2\times 2$ matrix. Such model has been analyzed in incommensurate magnetic orders in the diamond structure \cite{Lee2008-so} and the fcc lattice as a part of the diamond structure \cite{Balla2020-jp}. 
We focus on the orders with the ordering wavenumber at the L points: $\bk_1\equiv (\pi,\pi,\pi)$, $\bk_2\equiv (-\pi,\pi,\pi)$, $\bk_3\equiv (-\pi,-\pi,\pi)$, and $\bk_4\equiv (\pi,-\pi,\pi)$ as shown in Fig.~\ref{fig:1}(a). Since the exchange interactions at the L points are isotropic due to the symmetry, we retain such exchange interactions while neglecting  anisotropic ones in the following. Although the anisotropic ones  are generally present and can influence the ordering vectors and the excitation energy of quadrupole wave, they have no effect on the transition temperature for the order at the L points. Thus, a model without anisotropic interactions can be a good starting model as a minimal one.  
In this section, we consider  exchange interactions up to the fourth neighbor interactions, while we will simplify the model in the later sections. 

The $2\times 2$ blocks in Eq.~(\ref{eq:H}) are given by
\begin{align}
	{\mathcal H}^{AA}_{\bk}&=[J_2\gamma_2(\bk)+J_4\gamma_4(\bk)]
	\hat{1},\\
	{\mathcal H}^{AB}_{\bk}&=\frac{1}{2}[J_1\gamma_1(\bk) +J_3\gamma_3(\bk)]\hat{1}.
\end{align}
Here, $\hat{1}$ is a 2$\times$2 identity matrix.  $J_{i}$'s are isotropic coupling constants between the $i$th neighbor sites as indicated in Fig.~\ref{fig:0}(a). The expressions for $\gamma$'s are listed in Table \ref{table:tbl1}. Note that $\gamma_{1}(\bk)$ and $\gamma_3(\bk)$ are complex. From the neutron experimental data, the ordering patterns are consistent with those expected for $J_1>0$ \cite{Iwasa2017-ni}. 

For describing the symmetry breaking at the L-point modulation within our model, it is important to identify relevant interactions. As will be explained later, four special wavenumbers at the $\Gamma$ point and the X points $\{(2\pi,0,0), (0,2\pi,0), (0,0,2\pi)\}$ are important other than the L point. First, among the same sub-lattice interactions, we note that the next-nearest interaction $\gamma_2$ vanishes at the four L points, while the fourth neighbor one $\gamma_4$ is finite as shown in Table~\ref{table:tbl1}. This means that positive $J_4$ favors the order at the L points while does not favor those at the $\Gamma$ and X points. For the inter sublattice ones, the nearest-neighbor interaction $\gamma_1$ and the third-neighbor one $\gamma_3$ are finite, while both vanish at the X points. In this sense, the effects of $J_1$ and $J_3$ are similar. 
Diagonalizing the matrix in Eq.~(\ref{eq:H}), one obtains the eigenvalues $J(\bk)$. Figures \ref{fig:1}(b)--\ref{fig:1}(d) show $J(\bk)$ for typical parameter sets with $J_1=1$ as a unit of energy. For the nearest-neighbor model with $J_4=0$ in Fig.~\ref{fig:1}(b),  $J(\bk)$ is smallest at the $\Gamma$ point, leading to a N\'eel type antiferoic  quadrupole order. 
When $J_4$ is sufficiently large or $J_2$ is positive and not too large, the $J(\bk)$ has minima at the L points as shown in Figs.~\ref{fig:1}(c) and \ref{fig:1}(d), and these cases are situations we analyze in this paper.  For $J_1$-$J_2$ model, it is known that the frustration leads to various incommensurate orders in the diamond lattice Heisenberg magnet \cite{Bergman2007}. Since there is no experimental information about the exchange coupling constants, we try to understand the basic phase diagram for the quadrupole orders in PrIr$_2$Zn$_{20}$ \cite{Onimaru2011-of,Ishii2011-mj,Kittaka2024-cn}, focusing on the parameter sets that lead to the orders at the L points.

\begin{table*}[ht]
  \caption{exchange interaction coefficients. The values of $\gamma(\bk)$'s at $\bk=(0,0,0),(0,0,2\pi)$, $\bk_1=(\pi,\pi,\pi)$, and the other L points are also shown. Here, we take the unit cell convention of $A$: (0,0,0) and $B$: $(\tfrac{1}{4},\tfrac{1}{4},\tfrac{1}{4})$ and thus, the $\bk_1$ is inequivalent to the remaining $\bk_{2,3,4}$. $\eta_{\bk}\equiv e^{i(k_x+k_y+k_z)/4}$, $\zeta^\pm_{\bk}\equiv \pm 2(\cos\frac{k_x}{2}+\cos\frac{k_y}{2}+\cos\frac{k_z}{2})-3$. The mean-field second-order transition temperatures $T_c$'s are $T_c=2|J_1-3J_3|+6J_4$ for the L points, $T_c=4|J_1+3J_3|-12J_2-6J_4$ for the $\Gamma$ point, and $T_c=4J_2-6J_4$ for the X points.}
  \label{table:tbl1}
  \centering
  \begin{tabular}{cccccc}
notation & expression &$\Gamma(0,0,0)$ &X(0,0,$2\pi$)&L($\pi,\pi,\pi$) & L(others)\\
  \hline
  \hline\\[-2mm]
	$\gamma_2(\bk)$ & $2\left(\cos \frac{k_x}{2}\cos\frac{k_y}{2}+\cos \frac{k_y}{2}\cos \frac{k_z}{2}+\cos\frac{k_z}{2}\cos\frac{k_x}{2}\right)$   &$6$&$-2$&$0$ & $0$\\
	$\gamma_4(\bk)$ & $\cos k_x+\cos k_y + \cos k_z$ &$3$&$3$&$-3$ & $-3$\\[1mm]
	\hline\\[-2mm]
$\gamma_1(\bk)$ & $4\eta_{\bk}\left(\cos\frac{k_x}{4}\cos\frac{k_y}{4}\cos\frac{k_z}{4}+i\sin\frac{k_x}{4}\sin\frac{k_y}{4}\sin\frac{k_z}{4}\right)$ & $4$ & $0$ & $-2$ & $2$\\[2mm]
$\gamma_3(\bk)$ & $4\eta_{\bk}\left(
\cos\frac{k_x}{4}\cos\frac{k_y}{4}\cos\frac{k_z}{4}\zeta^+_{\bk}+i\sin\frac{k_x}{4}\sin\frac{k_y}{4}\sin\frac{k_z}{4}\zeta^-_{\bk}\right)$ & $12$ & $0$ & $6$ & $-6$\\[2mm]
\hline
\hline
  \end{tabular}
\end{table*}

\subsection{Landau Expansion}\label{landau}
We focus on the order parameter at the four equivalent L points. Note that $\bk_1+\bk_2+\bk_3+\bk_4\equiv {\bf 0}$ and that the fields at the L points are real since $\bk_n\equiv -\bk_n$.

The Landau's free energy ${\mathcal F}$ for the quadrupole moments is written as
\begin{align}
	{\mathcal F}&=\sum_{\bk}
	[\QQ^A_{-\bk},
	\QQ^B_{-\bk}]
	 \left[\frac{T}{2}+\hat{J}(\bk)\right]
	 \begin{bmatrix}
	 \QQ^A_{\bk}	 	\\
	 \QQ^B_{\bk}
	 \end{bmatrix}
	 \nonumber\\
	&+\sum_\alpha\Big[-b \sum_{\bk,\bk',\bk''}u^\alpha_{\bk} \Big(u^\alpha_{\bk'}u^\alpha_{\bk''}-3v^\alpha_{\bk'}v^\alpha_{\bk''}\Big)	\delta_{\bk+\bk'+\bk'',{\bm G}}\nonumber\\
	&+\frac{c}{4}\sum_{\bk,\bk',\bk'',\bk'''}(\QQ^\alpha_{\bk}\cdot \QQ^\alpha_{\bk'})(\QQ^\alpha_{\bk''}\cdot \QQ^\alpha_{\bk'''})\delta_{\bk+\bk'+\bk''+\bk''',{\bm G}}\Big].\label{eq:Forig}
\end{align}
Here, $\hat{J}(\bk)$ is the $4 \times 4$ matrix in Eq.~(\ref{eq:H}) with its eigenvalue $J(\bk)$, $T$ is the temperature, and $\bm{G}$ is a reciprocal lattice vector. We have taken into account local cubic and fourth-order terms only, which arise from local anisotropy such as CEF potential. We note that the cubic term $b$ arises from the effects of the excited states and relatively smaller than $c$ as estimated by using the actual CEF scheme \cite{Hattori2014}.

\subsection{L point free energy}
We concentrate on the L point components $\QQ^\alpha_n=(u_n^\alpha,v_n^\alpha)\equiv \QQ^\alpha_{\bk_n} (n=1,2,3,4)$ with the short-handed notation, we find the part of the free energy consisting of only $\QQ^\alpha_{\bk_n}$'s is 
\begin{align}
{\mathcal F}_{\rm L} &= \frac{a_{\rm L}}{2}\sum_{\alpha,n} |\QQ_n^\alpha|^2 + J_{\rm L}\sum_n (-1)^{p_n}\QQ^A_n \cdot \QQ^B_n
	+ \frac{c}{4}\sum_{\alpha,n} |\QQ_n^\alpha|^4\nn\\
	&+\frac{c'}{4}\sum_\alpha \sum_{m<n}|\QQ^\alpha_m|^2|\QQ^\alpha_n|^2+\frac{c''}{4}\sum_\alpha \sum_{m<n}
	(\QQ_m^\alpha\cdot \QQ_n^\alpha)^2\nonumber\\
	&+\frac{c'''}{4}\sum_\alpha \sum_{(m,n,m',n')}(\QQ^\alpha_m\cdot \QQ^\alpha_n)(\QQ^\alpha_{m'}\cdot \QQ^\alpha_{n'}).\label{eq:FLtot}
\end{align}
Note that $b$ terms are absent owing to the wavenumber conservation and  $(m,n,m',n')=(1,2,3,4)$, $(1,3,2,4)$, and $(1,4,2,3)$ in the last term in the right-hand side. The factor $p_n(n=1,2,3,4)$ represents the sign of exchange interaction between the $A$ and $B$ sublattices for four types of L points. Since our unit cell consists of the bond parallel to $\bk_1$, $p_n=0$ is distinct from the others $p_{2,3,4}=1$.  
 The coupling constants are denoted as 
\begin{align}
a_{\rm L}&\equiv T-6J_4, \quad J_{\rm L}\equiv -2J_1+6J_3,\\ & c'=2c, \quad c''=4c,\quad c'''=8c.
\label{eq:const_c}	
\end{align}
Note that $\bk_n\equiv -\bk_n$ and thus the $\QQ_n \cdot \QQ_n$ pair in the summation in $\sum_{\bk}$ appears once. 
Since the three fourth-order terms are independently allowed by the cubic symmetry, we denote them as independent parameters.

For discussing a single-$\bq$ order at the L points, $c'$, $c''$, and $c'''$ terms are irrelevant and the fourth-order free energy is simply given by 
\begin{equation}
	{\mathcal F}_L^{1\bq(4)}=\frac{c}{4}\sum_\alpha|\QQ^\alpha|^4.
\end{equation}
When multiple-$\bq$ states are considered, the coupling constants $c'$, $c''$, and $c'''$ matter as discussed in Sec.~\ref{sec:dis}.

\subsection{$\Gamma$ point free energy}
Under finite magnetic fields, the uniform quadrupole moment $\QQ_0^\alpha\equiv\QQ_{\bk=\bf 0}^\alpha$ is induced via the effective field $\tilde{\bm{h}}$ in the second-order perturbation of the magnetic field $\bm{H}$. The free energy up to the fourth-order  terms for the $\Gamma$-point modes are 

\begin{align}
	{\mathcal F}_\Gamma &=\sum_\alpha \Big(-
	\tilde{\bm h}\cdot  \QQ_{0}^{\alpha}
+\frac{a_{\Gamma 0}}{2} |\QQ_{0}^\alpha|^2\Big)
	+J_{\Gamma 0}\QQ_{0}^{A} \cdot \QQ_{0}^{B}\nn\\
	&+\sum_\alpha\Big\{-b u_0^\alpha\Big[\big( u_0^\alpha \big)^2 -3\big(v_0^\alpha\big)^2\Big]+\frac{c}{4}|\QQ_0^\alpha|^4 \Big\},\label{eq:F2Geff}
\end{align} 
where 
\begin{align}
	a_{\Gamma 0}&=T+12J_2+6J_4,\quad 
	J_{\Gamma 0}=4J_1+12J_3.
\end{align}
The effective field is given by 
\begin{align}
	\tilde{\bm{h}}=
	\begin{bmatrix}
		\tilde{h}_u\\
		\tilde{h}_v
	\end{bmatrix}\equiv g
	\begin{bmatrix}
		2H_z^2-H_x^2-H_y^2\\
		\sqrt{3}(H_x^2-H_y^2)
	\end{bmatrix},\label{eq:magH}
\end{align}
 and $g=(g_J\mu_{\rm B})^2[7/(3E_4)-1/E_5]$ with the excited 
$\Gamma_{4,5}$ CEF energy $E_{4,5}$ \cite{Hattori2014}. $\mu_{\rm B}$ is the Bohr magneton and $g_J$ is the Lande's $g$ factor for the multiplet with the angular momentum $J$; $g_J=4/5$ for the PrIr$_2$Zn$_{20}$ with $J=4$.

\subsection{Mode coupling terms}\label{modecoupling}
We now discuss mode coupling terms consisting of the fields at the L points and those at the $\Gamma$ point. The lowest-order contributions come from the cubic terms as
\begin{align}
	\delta {\mathcal F}_3=& -3\lambda_b\sum_{\alpha,n} \Big\{u_0^\alpha\Big[ \big(u_n^\alpha\big)^2 -\big(v_n^\alpha\big)^2 \Big]-v_0^\alpha (2u_n^\alpha v_n^\alpha)\Big\}\nn\\
	&=-3\lambda_b \sum_{\alpha,n}\phi_0^\alpha \big(\phi_{n}^\alpha\big)^2\cos(\theta^\alpha_0+2\theta^\alpha_n)\label{eq:deltaF3}.
\end{align}
Here, we have parameterized
\begin{align}
\QQ^\alpha_{\bf 0}&=\phi^\alpha_0 \begin{bmatrix}
	\cos\theta^\alpha_0\\
	\sin\theta^\alpha_0	
\end{bmatrix},\quad 
\QQ^\alpha_n=\phi^\alpha_{n} \begin{bmatrix}
	\cos\theta^\alpha_{n}\\
	\sin\theta^\alpha_{n}	
\end{bmatrix},
\end{align}
and the coupling constant $\lambda_b$ has been introduced as an independent parameter. If $\lambda_b$ is estimated by Eq.~(\ref{eq:Forig}), $\lambda_b=b$. 
The next leading order terms are in the fourth order and given by
\begin{align}
	\delta {\mathcal F}_4= \sum_{\alpha ,n} \left[ \frac{\lambda^{(1)}_c}{2}|\QQ_n^\alpha|^2|\QQ_0^\alpha|^2+\lambda_c^{(2)} (\QQ_n^\alpha\cdot\QQ_0^\alpha)^2\right]. \label{eq:deltaF4}
\end{align}
Again, the coupling constant $\lambda_c^{(1)}$ and $\lambda_c^{(2)}$ have been introduced as independent parameters. If $\lambda_c^{(1,2)}$ is estimated by Eq.~(\ref{eq:Forig}), $\lambda^{(1)}_c=\lambda_c^{(2)}=c$. 

In principle, $\QQ_0^\alpha$ can couple with $\QQ_{\bk}^\alpha$ in the form of $\sim \QQ_0^\alpha \QQ_\bk^\alpha\QQ_{-\bk}^\alpha$ and $\QQ_0^\alpha \QQ_0^\alpha \QQ_\bk^\alpha\QQ_{-\bk}^\alpha$. In this paper, we retain the order parameter $\QQ_n^\alpha$ and the induced moment $\QQ_0^\alpha$ under magnetic fields since the other fields are not important for the phase transition with the L point fields. Some renormalization effects arising from fields at the X points are discussed in Sec.~\ref{sec:MultipleQ} and Appendix~\ref{app:reno}.

\section{Results: single-q orders}\label{sec:results}
In this section, we carry out minimization of the free energy ${\mathcal F}_L+{\mathcal F}_\Gamma+\delta {\mathcal F}_3+\delta {\mathcal F}_4$ [Eqs.~(\ref{eq:FLtot}), (\ref{eq:F2Geff}), (\ref{eq:deltaF3}),  and (\ref{eq:deltaF4})],  
assuming the single-$\bq$ orders. The purpose here is to understand how mode coupling effects affect the order parameter configuration. The results in this section will serve as the basis for understanding the full phase diagram shown in Sec.~\ref{sec:MultipleQ}, where we will take into account multiple-$\bq$ configurations. 
Since we are interested in the phase transition owing to the order parameters at the L point, we assume the coefficients of the quadratic terms in the free energy at $\bk\ne\bk_n$ are always positive. We first discuss the candidates of the stable configurations by analyzing the third- and fourth-order terms in ${\mathcal F}_{\rm L}$. Then, we will show the numerical results of full temperature-field phase diagram.

\subsection{Qualitative analysis}\label{sec:IIIA}

First, we notice that the inter-sublattice correlations are governed by $J_{\rm L}=-2J_1+6J_3$ for $\QQ_n^\alpha$ in Eq.~(\ref{eq:FLtot}). To reproduce the quadrupole configuration propposed by the neutron scattering experiments \cite{Iwasa2017-ni}, $J_{\rm L}<0$ is required. Thus, it is legitimate to set 
\begin{align}
	&\QQ_{\rm L}\equiv \QQ_1^A=\QQ_1^B=
	\begin{bmatrix}
		u_{\rm L}\\
		v_{\rm L}
	\end{bmatrix}=
	\phi_{\rm L}\begin{bmatrix}
		\cos\theta_{\rm L}\\
		\sin\theta_{\rm L}
	\end{bmatrix}.
\end{align}
Here, we have chosen one of the L points at $\bk_1$ as a representative one since the other three fields at $\bk_{n} (n=2,3,4)$ with $\QQ_n^A=-\QQ_n^B$ are equivalent and this is also the case even when $\bm{H}$ is finite. See Table~\ref{table:tbl1}. For the modes at the $\Gamma$ point, which are gapped and just induced by magnetic fields or by the square of the order parameter at the L point, we set 
\begin{align}
	&\QQ_0\equiv \QQ_0^A=\QQ_0^B
	=\begin{bmatrix}
		u_0\\
		v_0
	\end{bmatrix}=
\phi_0\begin{bmatrix}
		\cos\theta_0\\
		\sin\theta_0
	\end{bmatrix}.
\end{align}

\begin{figure}[t!]
\begin{center}
\includegraphics[width=0.5\textwidth]{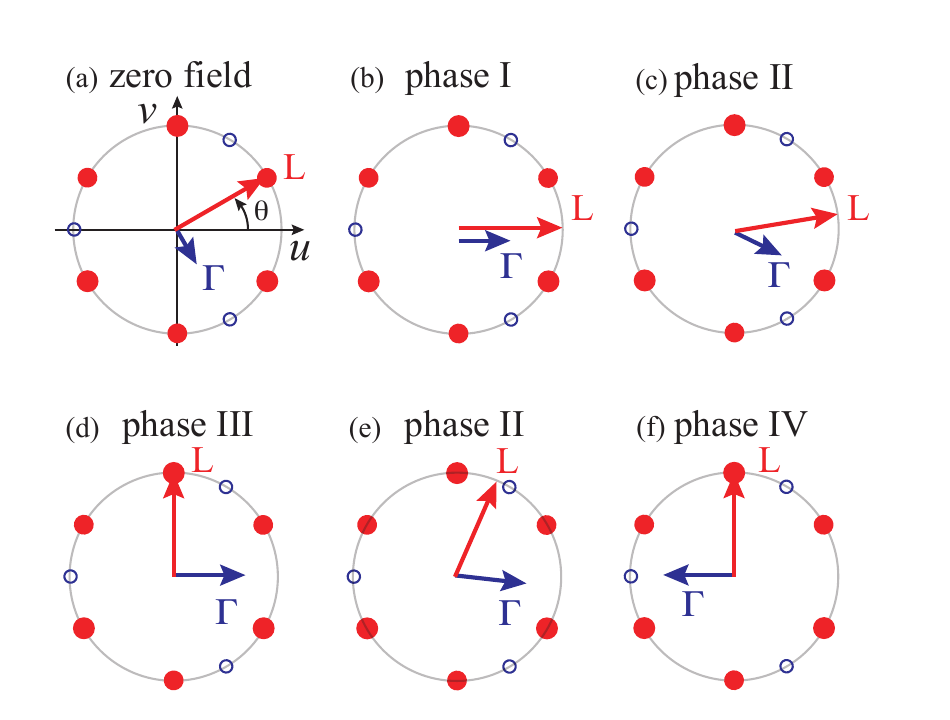}
\end{center}

\caption{Schematic single-$\bq$ order parameter configurations for (a) ${\bm H=0}$, (b) small ${\bm H}$, (c) small ${\bm H}$ and low $T$, and (d) large ${\bm H}$.  Blue (red) arrows indicate $\QQ_0$($\QQ_{\rm L}$). Filled red circles represent the directions $\theta_{\rm L}$ given by Eq.~(\ref{eq:6domain}) and open blue circles do $\theta_{\rm 0}$. A configuration for large ${\bm H}$ in the phase II is shown in  (e) and that for ${\bm H}\parallel [110]$ in (f). 
}
\label{fig:config}
\end{figure}

Second, we note that the single-$\bq$ free energy ${\mathcal F}_{\rm L}$ is isotropic. This means that the transition temperatures for $u_{\rm L}$ and $v_{\rm L}$ are the same at ${\bm H}={\bf 0}$ and $T_c=6J_4+2|J_1-3J_3|$ . This isotropic nature, however, is destroyed by the mode coupling to the fields at the $\Gamma$ point through $\delta {\mathcal F}_3$ and $\delta {\mathcal F}_4$.
When $T\lesssim T_c$, the induced $\QQ_0$ is sufficiently small, and thus, the third- and fourth-order terms in ${\mathcal F}_\Gamma$ can be irrelevant. In this situation the relative directions between $\QQ_n$ and $\QQ_0$ is determined by $\delta {\mathcal F}_3$ and $\delta {\mathcal F}_4$ as 
\begin{align}
	\cos(\theta_0+2\theta_{\rm L})=1 \ {\rm and}\ \cos(\theta_0-\theta_{\rm L})=0. \label{eq:F34mincond}
\end{align}
This leads to 
\begin{align}
	\theta_0=\frac{(2j-1)\pi}{3},
	\quad \theta_{\rm L}=\frac{(2j-1)\pi}{3}\pm\frac{\pi}{2}, \label{eq:6domain}
\end{align}
with $j=1,2,3$. These six domains are degenerate at 
${ \bm H}={\bf 0} \ (\tilde{\bm h}={\bf 0})$. See Fig.~\ref{fig:config}(a).
We should comment that the three directions of $\QQ_0$ do not coincide with those favored by the $b (>0)$ term in ${\mathcal F}_\Gamma: -b\phi_0^3\cos(3\theta_0)$. Since $\QQ_0$ is not the primary order parameter, the third-order term is expected to play a minor role and the configurations (\ref{eq:6domain}) are robust at least $\tilde{\bm h}={\bf 0}$.


\begin{figure*}[t!]
\begin{center}
\includegraphics[width=\textwidth]{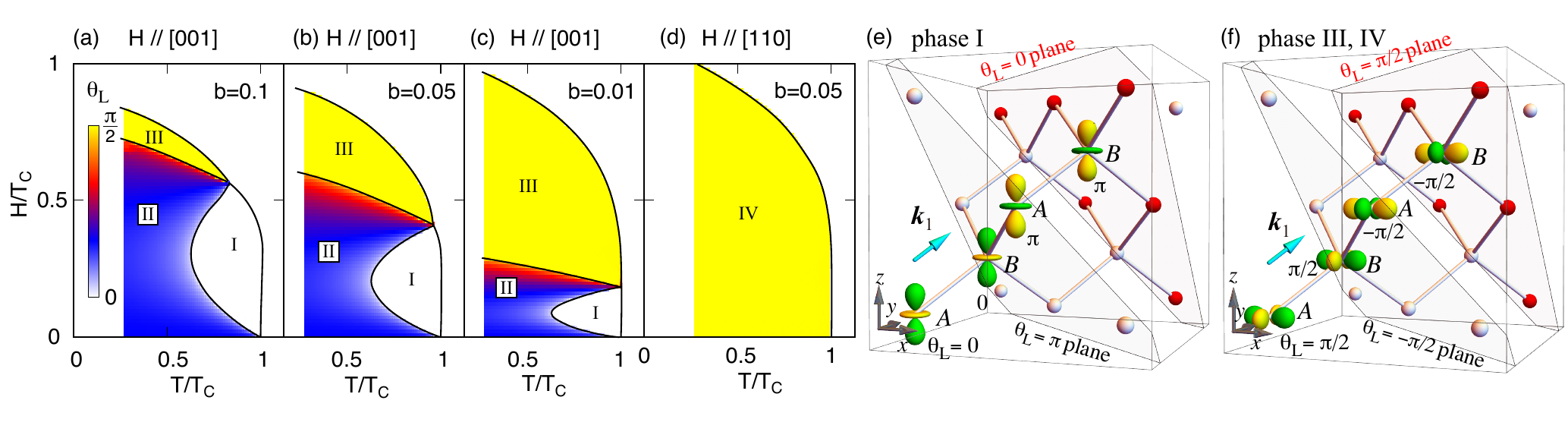}
\end{center}
\caption{(a)--(c) $T$--$H$ phase diagram for $\bm{H}\parallel [001]$, $J_1=1.0$, $J_2=J_3=0.0$, $J_4=1.0$, and $c=0.5$. (a) $b=0.1$, (b) $b=0.05$, and (c) $b=0.01$. Color map represents the L point order parameter angle $\theta_{\rm L}$. (d) $T$--$H$ phase diagram for $\bm{H}\parallel [110]$ with the same parameter set as in (b). The schematic L point order parameter configuration for (e) the phase I and for (f) the phases III and IV. For clarity, orbital shapes are drawn only for four sites. Red (white) spheres represent Pr sites with $\theta_{\rm L}=0\ (\pi)$ for (e) and  $\theta_{\rm L}=\pi/2\ (-\pi/2)$ for (f), respectively. On shaded (111) planes, the order parameter $\boldsymbol{\phi}^A(\bm{r})$ is uniform and $\boldsymbol{\phi}^B(\br)=\boldsymbol{\phi}^A(\br)$. For the phase II, the L point order parameter is in between those in the phase I and III with $0<|\theta_{\rm L}|<\pi/2$. }
\label{fig:phase}
\end{figure*}

Now let us consider the cases for finite magnetic fields. For high symmetric directions $\bm{H}=\tfrac{H}{\sqrt{2}}(\sin\Theta,\sin\Theta,\sqrt{2}\cos\Theta)$ with $0\le \Theta\le \pi/2$, the effective field reads as 
$\tilde{h}_u=gH^2(3\cos^2\Theta-1)$ and $\tilde{h}_v=0$. Note that $\tilde{\bm h}={\bf 0}$ at $\bm{H}\parallel [111]$ and $\tilde{h}_u>0$ for $\Theta<\tan^{-1}\sqrt{2}\equiv \Theta_{111}$, while $\tilde{h}_u<0$ for $\Theta>\Theta_{111}$. The effective field induces $u_0$, which leads to two ``transition temperatures'' $\tilde{T}_c^u$ for the $u_{\rm L}$ order and $\tilde{T}_c^v$ for $v_{\rm L}$. They are given by  
\begin{align}
	\tilde{T}_c^u&=T_c+6\lambda_b u_0 -3\lambda_cu_0^2, \ 
	\tilde{T}_c^v=T_c-6\lambda_bu_0 -\lambda_cu_0^2,
\end{align}
where we have set $\lambda_c^{(1)}=\lambda_c^{(2)}=\lambda_c$ for simplicity. 
Note that $\tilde{T}_c^{u,v}$ is a hypothetical transition temperature from the disordered state. The difference between the two is $\Delta \tilde{T}_c=\tilde{T}_c^u-\tilde{T}_c^v=2\lambda_c(\phi_0^*-u_0)u_0$ with $\phi_0^*\equiv 6\lambda_b/\lambda_c$. For $u_0>0$, this means $\tilde{T}_c^u>\tilde{T}_c^v$ for $u_0<\phi_{0}^*$, i.e., for low fields, while $\tilde{T}_c^v>\tilde{T}_c^u$ for the higher field with $u_0>\phi_{0}^*$. In contrast, $\Delta \tilde{T}_c<0$ for $u_0<0$. 

As is clear from the expression of 
$\Delta \tilde{T}_c$, it is expected 
that there is no additional phase 
transition within the single-$\bq$ 
$\QQ_n$ sector for $u_0<0$, i.e., 
$\Theta>\Theta_{111}$. 
In the following, we will 
concentrate on the cases for 
$u_0>0$ ($\Theta<\Theta_{111}$). 
For $u_0>0$, the linear coupling between 
$\QQ_0$ and $\tilde{h}_u$ can alter 
the stable configuration from the 
above six domains (\ref{eq:6domain}) 
since their energy 
gain is given by the second order in 
$\QQ_0$. 
Since $\Delta\tilde{T}_c>0$, the phase transition from the disordered phase occurs in the $u_{\rm L}$ component: $\theta_{\rm L}=0$ or $\pi$. Here, the two choices $0$ and $\pi$ represent a simple translation and we will consider $\theta_{\rm L}=0$ below. 
For small $\tilde{h}_u$, $\delta {\mathcal F}_3$ is more important than $\delta {\mathcal F}_4$. This means $\theta_0=0$ satisfying $2\theta_{\rm L}+\theta_0=0$ [Fig.~\ref{fig:config}(b)]. Such a 
collinear state possesses the free 
energy ${\mathcal F}_{\rm col}$ and the 
fluctuations around this is given as 
 \begin{align}
 	\delta {\mathcal F}_{\rm col}&\simeq \lambda_c\phi_0\phi_{\rm L}^2
 	 \begin{bmatrix}
 	 	\theta_0 & \theta_{\rm L}
 	 \end{bmatrix}\hat{A}	  	 \begin{bmatrix}
 	 	\theta_0\\
 	 	\theta_{\rm L}
 	 \end{bmatrix},\label{eq:hess0}\\
 	 \hat{A}&=
\begin{bmatrix}
 	 	(\lambda_c\phi_{\rm L}^2)^{-1}(\tilde{h}_u+9b\phi_0^2) +\frac{1}{2}\phi_0^*-2\phi_0 & \phi_0^*+2\phi_0\\
 	 	\phi_0^*+2\phi_0 & 2\phi_0^*-2\phi_0
 	 \end{bmatrix}. 	 
 	 \label{eq:hess}
 	 \end{align}
Indeed, the instability to finite $\theta_{{\rm L},0}$ occurs when one of the eigenvalue of the matrix $\hat{A}$ in Eq.~(\ref{eq:hess}) becomes zero. This takes place when  
\begin{align}
	\tilde{h}_u^{c1}=27\lambda_b\frac{\phi_0}{\phi_0^*-\phi_0}\phi_{\rm L}^2-9b\phi_0^2.\label{eq:hc_u}
\end{align}
Note that the factor $\phi_0^*-\phi_0$ of the denominator coincides with the factor in $\Delta T_c$. The configuration with $\theta_0\ne 0$ and $\theta_{\rm L}\ne 0$ gradually crosses over to the configuration for $\bm H=0$. See Fig.~\ref{fig:config}(c).
When $\tilde{\bm h}$ is sufficiently large, $u_0$ grows and $\delta {\mathcal F}_{4}$ plays an important role. Thus, $\QQ_{\rm L}$ is locked at $\theta_{\rm L}=\pm \pi/2$ as shown in Fig.~\ref{fig:config}(d).  This is because $\delta {\mathcal F}_4>\delta {\mathcal F}_3$ for large $\tilde{h}_u$. See Eq.~(\ref{eq:F34mincond}).
By the similar analysis to Eqs.~(\ref{eq:hess0})--(\ref{eq:hc_u}), the boundary determined by the field below which $|\theta_{\rm L}|<\pi/2$ [Fig.~\ref{fig:config}(e)] is given as 
\begin{align}
	\tilde{h}_u^{c2}=27\lambda_b\frac{\phi_0}{\phi_0-\phi_0^*}\phi_{\rm L}^2-9b\phi_0^2.\label{eq:hc_v}
\end{align}
For sufficiently large $\lambda_b$, the $\theta_{\rm L}=\pm \pi/2$ phase never appears because of $\tilde{T}_c^v<0$. Of course, for extremely large $\tilde{\bm h}$, 
the L-point order vanishes trivially. 

Finally, we briefly comment about the cases for $u_0<0$ ($\Theta>\Theta_{111}$) under finite magnetic fields. As we have noted, there is no phase transition once a quadrupole moment with $\theta_{\rm L}=\pm \pi/2$ is ordered as shown in Fig.~\ref{fig:config}(f). This is because $\theta_{\rm L}=\pm \pi/2$ and $\theta_0=\pi$ can minimize $\delta {\mathcal F}_3$, $\delta {\mathcal F}_4$, and ${\mathcal F}_{\Gamma}$, simultaneously.

\subsection{Phase diagram}
Now we show the results of numerical minimization of the single-$\bq$ free energy (\ref{eq:FLtot}), (\ref{eq:F2Geff}), (\ref{eq:deltaF3}), and (\ref{eq:deltaF4}). We will discuss the result for ${\bm H} \parallel [001]$, i.e., $\tilde{h}_u>0$ and then for ${\bm H} \parallel [110]$, $\tilde{h}_u<0$. To understand the qualitative nature of the phase diagram, it is sufficient to discuss the cases for $J_2=J_3=0$.

Figures~\ref{fig:phase}(a)--\ref{fig:phase}(c) show the $T$--$H$ phase diagram for $\bm H \parallel [001]$, where $H=|\bm{H}|$ and the color represents the angle for the L point moment $\theta_{\rm L}$: (a) $b=0.1$, (b) $b=0.05$, and (c) $b=0.01$ with $c=0.5$ fixed. As discussed in the previous section, the phase III is stabilized under high magnetic fields. This phase, however, is suppressed when $b$ becomes larger. We note that the collinear phase (phase I) appears for high $T$ and intermediate $H$. The reason why the phase I cannot appear at the lower $T$ is understood by examining Eq.~(\ref{eq:hc_u}); there is no solution of the critical field $\tilde{h}_u^{c1}$ when $\phi_{\rm L}^2$, which is approximately $H$ independent, is large.

For $\bm{H} \parallel [110]$, the situation becomes much simpler. Figure \ref{fig:phase}(d) shows the $T$--$H$ phase diagram for $\bm{H} \parallel [110]$. As discussed in the last part in Sec.~\ref{sec:IIIA}, there is only one stable phase (phase IV), where $\theta_{\rm L}=\pm \pi/2$ and $\theta_0=\pi$. For $b=0.1$ and $b=0.01$, the phase diagrams are similar and we do not show them.  

Figures \ref{fig:phase}(e) and \ref{fig:phase}(f) show the real space quadrupole configurations with the ordering wave number $\bm{k}_1$ for 
$\bm{H} \parallel [001]$ and $\bm{H} \parallel [110]$, respectively. 
For both cases, the quadrupole orders with $\bm{\phi}_A=\bm{\phi}_B$ are ferroic on the (111) planes, while antiferroic between the neighboring (111) planes as shown in Fig.~\ref{fig:phase}. Namely the quadrupole moments at the sites indicated by white spheres and those by red ones have opposite sign.   When the ordering wave number is $\bm{k}_n (n=2,3,4)$, the quadrupole configurations are ferroic on the plane perpendicular to $\bk_n$ with $\bm{\phi}_A=-\bm{\phi}_B$.

\section{Results:  multiple-q orders} \label{sec:MultipleQ}
In this section, we consider multiple-$\bq$ orders for the quadrupole moments at the L points. We will first introduce a simple symmetric multiple-$\bq$ states with the same amplitudes for all the $\bq_n$ in Sec.~\ref{sec:symQ}. Then, we will show numerical results, where we relax the constraint about the amplitudes of the order parameters. 

\subsection{Free energy for symmetric states} \label{sec:symQ} 
We consider double-$\bq\   (\mathcal N=2)$ and quadruple-$\bq\  (\mathcal N=4)$ configurations, whose real space moments are given by 
\begin{equation}
	\QQ^\alpha(\br) =
		\begin{bmatrix}
		u^\alpha_{\br}\\
		v^\alpha_{\br}
	\end{bmatrix}=
	 \frac{\Phi_\alpha}{\sqrt{\mathcal{N}}}\sum_{n=1}^{\mathcal N} e^{i\bk_n\cdot \br}\begin{bmatrix}
		\cos\theta_n^\alpha\\
		\sin\theta_n^\alpha\\		
	\end{bmatrix},\label{eq:multiq_LC}
\end{equation}
where 
$\br$ is labeled by the $A$-sublattice position as before and we have assumed that the magnitude for each $\bk_n$ is the same $\Phi_\alpha/\sqrt{\mathcal N}~(>0)$ for simplicity (symmetric states). 
For $n=1$ $\bm{\phi}^A_1=\bm{\phi}^B_1$, while, for $n=2,3$ and $4$, $\bm{\phi}^A_n=-\bm{\phi}^B_n$. As mentioned before, this is due to our unit cell convention. In this normalization $(\mathcal N)$, the quadratic part is unchanged from the single-$\bq$ expression. Thus, types of order realized below the second-order transition at $T=T_c$ are determined by the fourth-order terms. 

For double-$\bq$ states, the fourth-order terms are
\begin{equation}
	{\mathcal F}_{\rm L}^{2\bq (4)}=\sum_\alpha\Phi_\alpha^4\left[\frac{c}{8}+\frac{c'}{16}+\frac{c''}{16}(c^\alpha_{12})^2\right]. \label{eq:Fdouble}
\end{equation}
Here, $c^\alpha_{mn}\equiv\cos(\theta^\alpha_{m}-\theta^\alpha_n)$. 
This clearly indicates that $\theta_1^\alpha-\theta_2^\alpha =\pm \pi/2$ for $c''>0$ (orthogonal double-$\bq$), while $\theta_1^\alpha-\theta_2^\alpha=0$ or $\pi$ for $c''<0$ (parallel double-$\bq$). As explained in Appendix~\ref{app:config}, the double-$\bq$ configurations consist of four ferroic chains running along the direction of $\bk_3+\bk_4$ when only the $A$-sublattice sites are concerned. 

For quadruple-$\bq$ states, 
\begin{align}
	{\mathcal F}_{\rm L}^{4\bq (4)}=&\sum_\alpha\Phi_\alpha^4\Bigg[\frac{c}{16}+\frac{3c'}{32}+\frac{c''}{64}\sum_{m<n}(c^\alpha_{mn})^2 \Bigg].\label{eq:Fquadruple}
\end{align}
Note that the $c'''$ terms cancel when summing up the terms for $\alpha=A$ and $B$, since  $\phi^A_n=(-1)^{p_n}\phi^B_n$. The minimum value of ${\mathcal F}_{\rm L}^{4\bq (4)}$ for $c''>0$ is realized with $\sum_{m<n}(c^\alpha_{mn})^2=2$, where two orthogonal pairs with the relative angle between the two is arbitrary, e.g., $(\theta^\alpha_1,\theta^\alpha_2,\theta^\alpha_3,\theta^\alpha_4)=(\theta^\alpha+\pi/2,\theta^\alpha,\varphi^\alpha+\pi/2,\varphi^\alpha)$. For $c''<0$, the configurations for the minimum free energy is, e.g., $(\theta^\alpha_1,\theta^\alpha_2,\theta^\alpha_3,\theta^\alpha_4)=(\theta,\theta,\theta,\theta)$ with $\sum_{m<n}(c^\alpha_{mn})^2=6$. The unit cell for the quadrupole-$\bq$ configurations contains eight $A$-sublattice sites with sixteen sites in total. See also Fig.~\ref{fig:app_2q4q} in Appendix~\ref{app:config}. 

Figure \ref{fig:phase2q4q} shows the phase diagram at $T=T_c$ on the $c''$--$c'$ plane for $c>0$. Note that there is no $c'''$ dependence. For large $c'$, multiple-$\bq$ states are not realized as is evident from Eqs.~(\ref{eq:Fdouble}) and (\ref{eq:Fquadruple}). The conditions given by Eq.~(\ref{eq:const_c}) just correspond to the point $(c'/c,c''/c)=(2,4)$, where  the single-$\bq$ and double-$\bq$ orders are degenerate. As long as the local fourth-order terms are dominant, quadruple-$\bq$ orders are unlikely. Thus, the $c'$ term is important when one compares the free energy for the single- and double-$\bq$ orders.


\begin{figure}[t!]
\begin{center}
\includegraphics[width=0.4\textwidth]{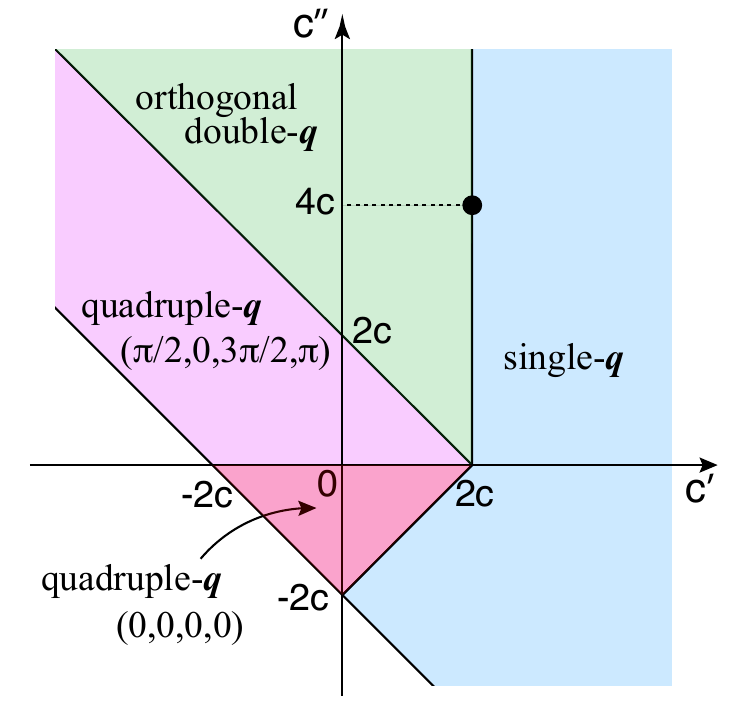}
\end{center}
\caption{Phase diagram at $T=T_c$ in the $c'$--$c''$ plane for $c>0$. A filled circle indicates the bare parameter set $c'=2c$ and $c''=4c$ given by Eq.~(\ref{eq:const_c}). Region for $c'+c''<-2c$ is unphysical since the fourth-order term becomes negative. For quadruple-$\bq$ states, their representative configurations for $(\theta^\alpha_1,\theta^\alpha_2,\theta^\alpha_3,\theta^\alpha_4)$ are indicated. }
\label{fig:phase2q4q}
\end{figure}

So far, we have treated the fourth-order coefficients $c$'s as phenomenological parameters. Here, we discuss the renormalization owing to modes other than $\bm{\phi}_0$ and $\bm{\phi}_n$. In the lowest orders, the cubic terms in Eq.~(\ref{eq:Forig}) are the most important. Among them, there are special terms which connect two L-point modes at $\bq_m$ and $\bq_n$ ($m\ne n$) via $\sim \phi_{\bm k}\phi_m \phi_n$, where $\bk$ is one of the wavenumbers at the X points $\bq_{\rm X}=(2\pi,0,0), \bq_{\rm Y}=(0,2\pi,0)$, or $\bq_{\rm Z}=(0,0,2\pi)$. The combination of $(\bk,\bq_m,\bq_n)$ is restricted by the Umklapp condition $\bk+\bq_m+\bq_n\equiv \bf{0}$ and there are two sets of $(m,n)$ for a given $\bk$. As shown in Appendix~\ref{app:reno}, these couplings lead to suppression of the coefficients $c'$ and $c'''$. See Eq.~(\ref{eq:app_reno}). Furthermore, the fifth-order couplings between the $\Gamma$- and the L-points fields appear after integrating out the X-point fields as $\sim \phi_0\phi_m^2\phi_n^2$ and $\sim \phi_0\phi_1\phi_2\phi_3\phi_4$. They play an important role under magnetic fields $\bm{H}$ since $\phi_0$ is induced by $\bm{H}$. Thus, these terms renormalize $c$'s with varying their magnitude depending on $\bm{H}$. 

The suppression of $c'$ and $c'''$ due to the renormalization with retaining $c''>0$ suggests that the orthogonal double-$\bq$ states are more stable. However, the coupling between the fields at the $\Gamma$ and the L 
points possesses an energy gain for the single-$\bq$ orders and double-$\bq$ ones with $\phi_m\ne\phi_n$. This is because there is a finite $\phi_0$ induced in the single-$\bq$ states, while there is no gain for the double-$\bq$ states with $\phi_n=\phi_m$ and $\theta_n-\theta_m=\pm \pi/2$. See the coupling in Eq.~(\ref{eq:deltaF3}). As we will see in Sec.~\ref{sec:2qphase}, the fifth-order terms generated by the integration of the X-points fields choose some particular $\theta_n$'s among degenerate orthogonal double-$\bq$ configurations.

\subsection{Phase diagram} \label{sec:2qphase}
\begin{figure}[t!]
\begin{center}
\includegraphics[width=0.48\textwidth]{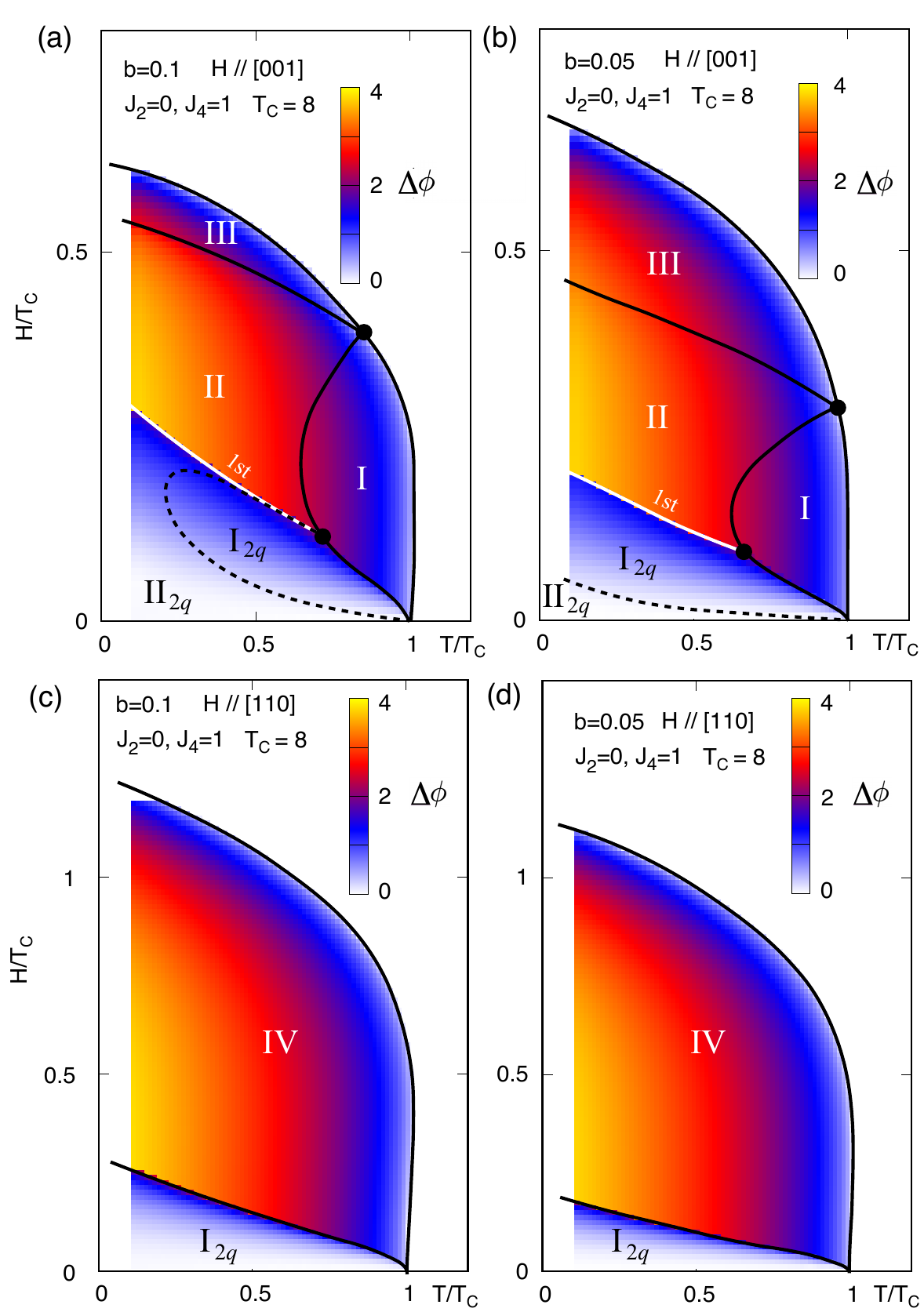}
\end{center}
\caption{$T$--$H$ phase diagram for $\bm{H}\parallel [001]$, $J_1=1.0$,  and $c=0.5$. (a) $(b,J_2,J_3,J_4)=(0.1,0.0,0.0,1.0)$, (b) $(0.05,0.0,0.0,1.0)$, and (c) $(0.03,0.5,0.3,0.4)$. Color represents $\Delta\phi\equiv|\phi_1-\phi_2|$, where double-$\bq$ orders with the ordering vectors $\bk_1$ and $\bk_2$ are assumed. For single-$\bq$ orders, we use the convention of $\phi_2=0$. Thick lines represent phase boundaries, where ``1st'' represents a line of first-order phase transition. Dashed lines are boundaries among double-$\bq$ configurations. Low $T$ regime is not shown since the Landau expansion is not valid in general. }
\label{fig:phase2q4q_simulation}
\end{figure}

Now, we discuss $T$--$H$ phase diagram ($H= |\bm{H}|$) taking into account multiple-$\bq$ orders in addition to the single-$\bq$ ones. We numerically minimize the free energy ${\mathcal F}_{\rm L}+{\mathcal F}_\Gamma+\delta {\mathcal F}_3+\delta {\mathcal F}_4 + \delta {\mathcal F}_{\rm {XL}}$, where $\delta {\mathcal F}_{\rm {XL}}$ is due to the integration of the X-point fields as discussed in Appendix \ref{app:reno} and the others are given in Eqs.~(\ref{eq:FLtot}),  (\ref{eq:F2Geff}), (\ref{eq:deltaF3}), and (\ref{eq:deltaF4}). To simplify the analysis, we assume $\QQ_n^B=\QQ_n^A(-1)^{p_n}$ valid for $J_{\rm L}<0$ and do not take into account quadruple-$\bq$ orders, since their energy is much higher than single- and double-$\bq$ states for the conventional fourth-order terms [Eq.~(\ref{eq:const_c})] as shown in Fig.~\ref{fig:phase2q4q}. 
Figures \ref{fig:phase2q4q_simulation}(a)--\ref{fig:phase2q4q_simulation}(d) show the 
phase diagrams for $\bm{H}\parallel [001]$, i.e., $\tilde{\bm{h}}=\tilde{h}_u(1,0)$ $(\tilde{h}_u\ge 0)$, 
 $(J_2,J_3,J_4)=(0.0,0.0,1.0)$ with (a) $b=0.1$ and (b) $b=0.05$. 
 For  $\bm{H}\parallel [110]$ with $\tilde{\bm{h}}=\tilde{h}_u(1,0)$ $(\tilde{h}_u\le 0)$, the corresponding phase diagrams are shown in Figs.~\ref{fig:phase2q4q_simulation}(c) and \ref{fig:phase2q4q_simulation}(d). One can notice that there are two double-$\bq$ phases labeled by I$_{2q}$ and II$_{2q}$, in addition to the single-$\bq$ phases I, II, III, and IV as already discussed above.


\begin{figure}[t!]
\begin{center}
\includegraphics[width=0.47\textwidth]{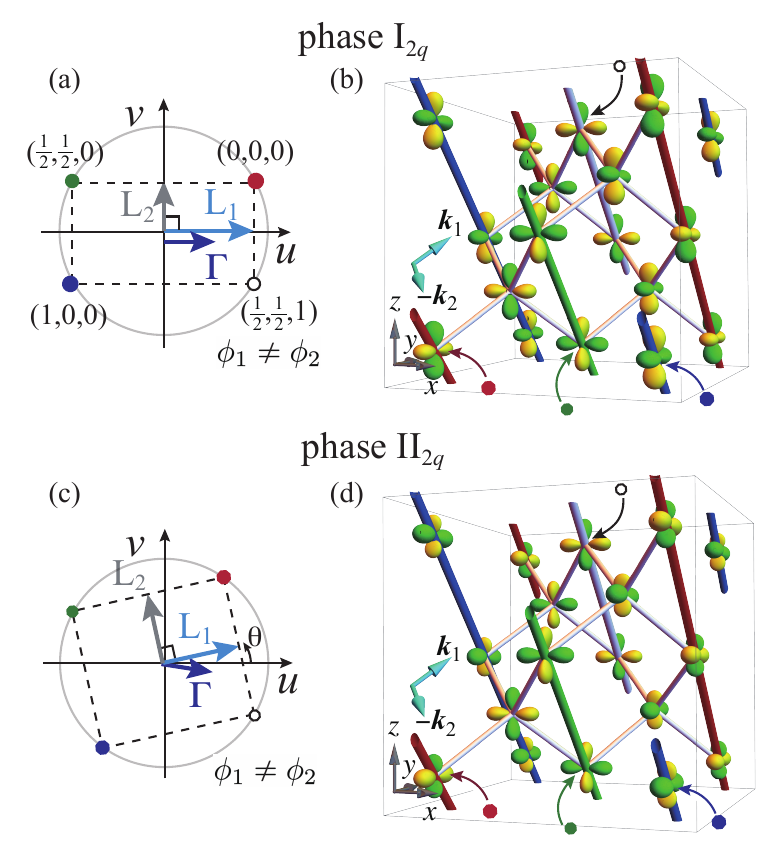}
\end{center}
\caption{Schematic orthogonal double-$\bq$ order parameter configurations for I$_{2q}$: $(\theta_1,\theta_2)=(0,\pi/2)$ with $\phi_1>\phi_2$ (a) in the $u$--$v$ space and (b) in the real space with $\phi_1=\sqrt{3}\phi_2$, and 
for II$_{2q}$:  (c) $(\theta_1,\theta_2)$$=$$(\theta,\theta+\pi/2)$ with  $\phi_1>\phi_2$ in the $u$--$v$ space and (d) $(\theta_1,\theta_2)$$=$$(\pi/24,13\pi/24)$ with $\phi_1=1.2\phi_2$ in the real space. In (a) and (c), arrows indicate the order parameters $\bm{\phi}_1$ (L$_1$), $\bm{\phi}_2$ (L$_2$), and $\bm{\phi}_0$ $(\Gamma)$. Four small circles represent the quadrupole moments at the $A$-sublattice real-space positions $\br=(0,0,0)$, $(1,0,0)$, $(\tfrac{1}{2},\frac{1}{2},0)$, and $(\tfrac{1}{2},\frac{1}{2},1)$.  
In (b) and (d), only L-point quadrupole components without those at the $\Gamma$ point are shown. Colored bonds running [0$\bar{1}1]$ directions connect the $A$-sublattice sites with the same value of $\boldsymbol{\phi}^A(\br)$. The colors of these bonds correspond to those of the circls in (a) and (c).}
\label{fig:config2q4q}
\end{figure}

Figure \ref{fig:phase2q4q_simulation}(a) shows the phase 
diagram  for $J_1=1$,  $c=0.5$, and $b=0.1$. The exchange coupling constants are the same as used in Fig.~\ref{fig:phase}: $(J_2,J_3,J_4)=(0.0,0.0,1.0)$. Except for the low-field region, the phase diagram is similar to those shown in Fig.~\ref{fig:phase}, showing the single-$\bq$ orders I, II, and III. An orthogonal double-$\bq$ configuration I$_{2q}$ with $\bm{\phi}_{\bk_1}=(\phi_1,0)$ and  $\bm{\phi}_{\bk_2}=(0,\phi_2)$ is stabilized for finite magnetic fields, touching its high-$T$ side boundary with the phase I.  The configuration of I$_{2q}$ state in the quadrupole space is shown in Fig.~\ref{fig:config2q4q}(a) and its real-space one is illustrated in Fig.~\ref{fig:config2q4q}(b), where a domain with the ordering wave numbers $\bk_1$ and $\bk_2$ is shown.
At $H=0$, the orthogonal double-$\bq$ states with $\phi_1=\phi_2$ discussed in Sec.~\ref{sec:symQ} are realized. As  $H$ increases, $\phi_1$ becomes larger than $\phi_2$. Since the phase I is the single-$\bq$ order with $v_{\rm L}=0$, the symmetry breaking from the phase I to the phase I$_{2q}$ is caused by the emergence of $\bm{\phi}_2=(0,\phi_2)$ through the second-order transition. In this sense, $\phi_1>\phi_2$ is quite natural. In the free energy, this is due to the $\delta {\mathcal F}_3$, which is minimized when the L point field is parallel to $\bm{\phi}_0=(u_0,0)$, while maximized when it is perpendicular. As for the phase boundary between the phase I$_{2q}$ and the phase II, it is of first-order, while the other phase boundaries are second-order.

There is another double-$\bq$ order labeled by II$_{2q}$
in Figs.~\ref{fig:phase2q4q_simulation}(a) and \ref{fig:phase2q4q_simulation}(b). The quadrupole configurations  in the phase II$_{2q}$ are ``rotated'' from those in the phase I$_{2q}$ with keeping $\bm{\phi}_1\perp \bm{\phi}_2$ as shown in Fig.~\ref{fig:config2q4q}(b). For smaller $b=0.05$ [Fig.~\ref{fig:phase2q4q}(b)], the phase I, II, and the double-$\bq$ orders shrink, while the phase III expands. The overall phase diagram, however, is unchanged.

For $\bm{H}\parallel [110]$, the phase I$_{2q}$ appears at 
low-field region as shown in Figs.~\ref{fig:phase2q4q_simulation}(c) and \ref{fig:phase2q4q_simulation}(d), while the phase II$_{2q}$ does not appear. In contrast to the phase I$_{2q}$ for $\bm{H}\parallel [001]$, the quadrupole configuration is $\bm{\phi}_1=(0,\phi_1)$ and $\bm{\phi}_2=(\phi_2,0)$ with $\phi_1>\phi_2$. This is natural since the neighboring single-$\bq$ phase is the phase IV. The fact that $\phi_1$ is larger than $\phi_2$ explains why there is no II$_{2q}$ for $\bm{H}\parallel [110]$. The coupling $\lambda_c^{(2)}$ in Eq.~(\ref{eq:deltaF4}) favors $\boldsymbol{\phi}_n \perp \boldsymbol{\phi}_0$. This suggests that the configuration in the phase I$_{2q}$ for $\bm{H}\parallel [110]$ is stable since the larger $\boldsymbol{\phi}_1$ is perpendicular to $\boldsymbol{\phi}_0$, while parallel in the phase I$_{2q}$ for $\bm{H}\parallel [110]$.


\begin{figure}[t!]
\begin{center}
\includegraphics[width=0.47\textwidth]{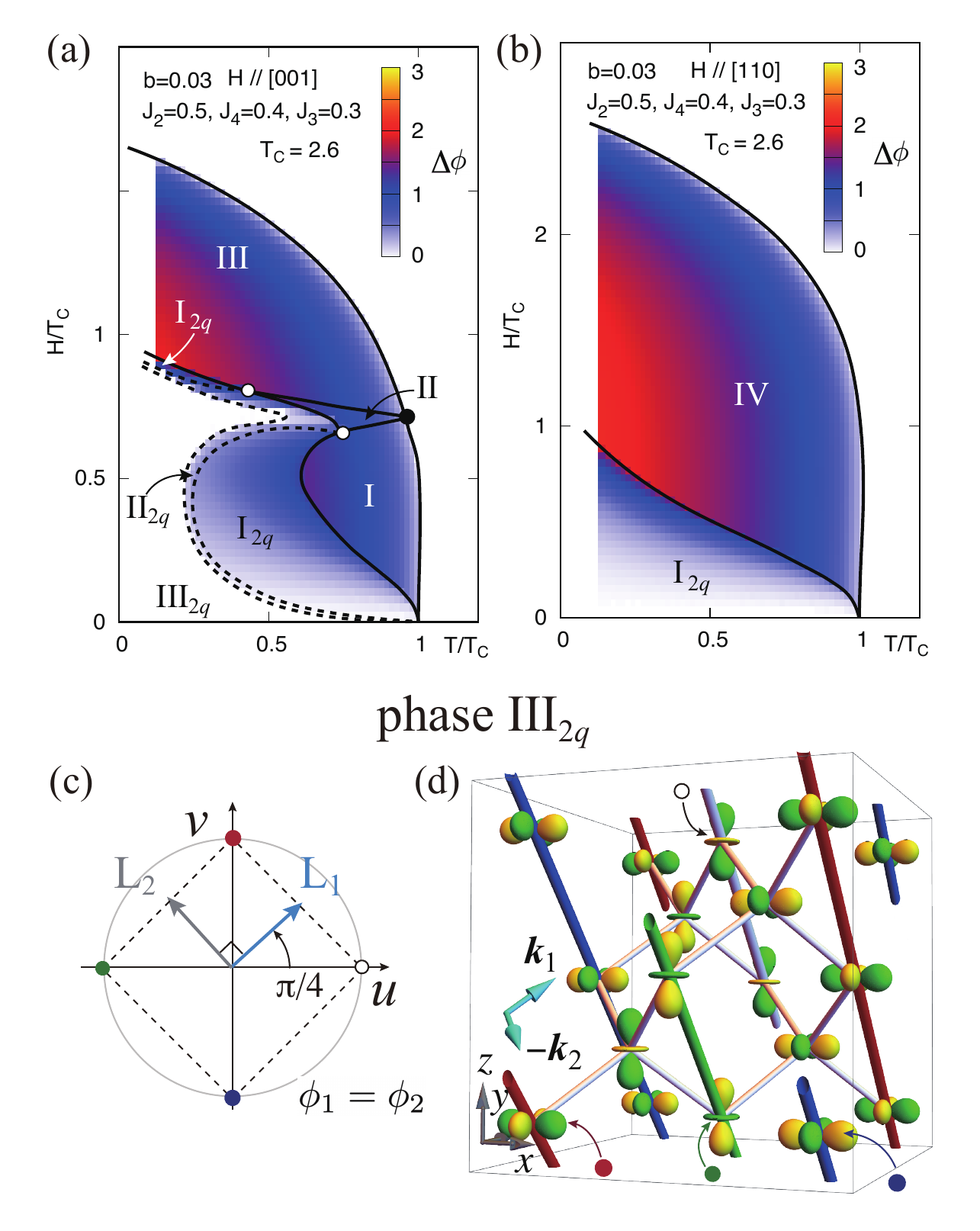}
\end{center}
\caption{$T$--$H$ phase diagram for $J_1=1.0$, $c=0.5$ and $(b,J_2,J_3,J_4)=(0.03,0.5,0.3,0.4)$ for (a) $\bm{H}\parallel [001]$ and (b) $\bm{H}\parallel [110]$.  
 Color represents $\Delta\phi\equiv|\phi_1-\phi_2|$, similarly to Fig.~\ref{fig:phase2q4q_simulation}. (c) Schematic orthogonal double-$\bq$ order parameter configurations in the phase III$_{2q}$ for $(\theta_1,\theta_2)=(\pi/4,3\pi/4)$ with $\phi_1=\phi_2$ (a) in the $u$--$v$ space and (b) in the real space. Colored bonds represent the uniform alignment of $\boldsymbol{\phi}^A(\br)$ similarly to those in Figs.~\ref{fig:config2q4q}(b) and \ref{fig:config2q4q}(d).
 }
\label{fig:phase2q4q_general}
\end{figure}

As an example of  more general cases, we show the phase diagram for $(b,J_2,J_3,J_4)=(0.03,0.5,0.3,0.4)$ in Fig.~\ref{fig:phase2q4q_general}. 
At low temperatures, the induced $\Gamma$-point moment $\bm{\phi}_0$ favors the configuration III$_{2q}$ shown in Figs.~\ref{fig:phase2q4q_general}(c) and \ref{fig:phase2q4q_general}(d),  where $\bm{\phi}_1=\tfrac{1}{\sqrt{2}}\phi_1(1,1)$ and $\bm{\phi}_2=\tfrac{1}{\sqrt{2}}\phi_2(-1,1)$ with $\phi_1=\phi_2$. 
This configuration is stabilized by the fifth-order coupling  discussed in the last part of Sec.~\ref{sec:symQ}. See also the last term in Eq.~(\ref{eq:appendixB1}). 

For this parameter set, the second-order coefficient $a_{\rm X}$ of the X-point fields  almost vanish at $T=0$ as shown in Fig.~\ref{fig:1}(d), the renormalization effects due to the X point fields grows rapidly at low temperatures. This causes the strong stabilization of the phase III$_{2q}$ in Fig.~\ref{fig:phase2q4q_general}(a) for $\bm{H}\parallel [001]$, showing reentrant behavior as a function of $H$. For $\bm{H}\parallel [110]$, the sharp increase of the critical field between I$_{2q}$ and IV phases in Fig.~\ref{fig:phase2q4q_general}(b) is also due to the fact that $a_{\rm X}$ is close to zero. Note that the Landau theory is not applicable to such low temperature region far away from $T\sim T_c$. There is a thin region of the phase II$_{2q}$ for $\bm{H}\parallel [001]$. The phase boundary between the phase II$_{2q}$ and II is of the second order. Similarly, that between the small region of phase I$_{2q}$ and III is also second order. We note that the boundaries indicated by the dashed lines are not symmetry breaking phase boundary since the symmetry of the three double-$\bq$ phases are the same as is evident from the real space configurations in Figs.~\ref{fig:config2q4q}(b), \ref{fig:config2q4q}(d), and \ref{fig:phase2q4q_general}(d). See also the discussion in Appendix~\ref{sec:appC}.

\section{Discussions}\label{sec:dis}

So far, we have shown how the L point quadrupole orders evolve under magnetic fields by using the Landau analysis. In this section, we will compare these results with the experimental data of PrIr$_2$Zn$_{20}$ and the related compounds. We also discuss the similarity between other quadrupole orders in Pr-based 1-2-20 compounds such as the N\'eel type antiferroic orders and those at the X points from theoretical point of view. In the last part of this section, we will comment on the future perspectives toward full understanding of the quadrupole orders in PrIr$_2$Zn$_{20}$. 

\subsection{Comparison with the experimental data of PrIr$_2$Zn$_{20}$}
PrIr$_2$Zn$_{20}$ shows several phase transitions under magnetic field at low temperatures as reported in the early study \cite{Onimaru2011-of}. However, the order parameters have not been identified except for the $x^2$$-$$y^2$ type quadrupole order with the ordering wave number at the L point for $\bm{H} \parallel [110]$ by the neutron scattering experiment \cite{Iwasa2017-ni}. The analysis of this study is based on this experimental fact. Namely, we have assumed that  the leading order parameter is the quadrupole moment at the L points. Constructing the symmetry-allowed free energy consisting of them and the related degrees of freedom, we have analyzed the phase diagram for $\bm{H} \parallel [001]$ in addition to that for $\bm{H} \parallel [110]$. Our analysis demonstrates that a $2z^2$$-$$x^2$$-$$y^2$ configuration with the ordering wave number at the L point is stable at high-$T$ and up to intermediate magnetic field strength for $\bm{H}\parallel [001]$. This suggests that the $2z^2$$-$$x^2$$-$$y^2$ order can be a promising candidate for the high-$T$ phase found for $\bm{H}\parallel [001]$ in the experimental works \cite{Onimaru2011-of,Kittaka2024-cn}. 
As for the order parameter at zero field, the single-$\bq$ configuration is not stable while the double-$\bq$ I$_{2q}$ is realized with $\phi_1=\phi_2$. For finite ${\bm H} \parallel [001]$, this double-$\bq$ configuration is modified and  $\bm{\phi}_1=(\phi_1,0)$ and $\bm{\phi}_2=(0,\phi_2)$ with $\phi_1>\phi_2$ is favored. $\phi_2$ increases as lowering $T$ and there appear phase II$_{2q}$ and III$_{2q}$ by simultaneously rotating $\bm{\phi}_1$ and $\bm{\phi}_2$. As the cubic anisotropy $b$ increases, the influence of the renormalized fourth-order coefficients becomes larger and favors the phase II$_{2q}$ and III$_{2q}$. 

As magnetic field increases, the double-$\bq$ order is destabilized and single-$\bq$ orders take place as discussed in Sec.~\ref{sec:results}. Thus, as increasing $\bm{H}\parallel [001]$, the double-$\bq$, the single-$\bq$ phase II with both $u_{\rm L}$ and $v_{\rm L}$ finite, and then the phase III with only $v_{\rm L}$ are realized. The trace of the phase boundaries between them has not been clearly observed in the experiments. However, there are several anomalies in the ultrasonic experiments \cite{Ishii2011-mj}. We hope the relation between our phase diagram and these anomalies can be clarified in future experimental studies. In addition to this, there is ambiguity about the stability of the phase III, whether the phase III extends toward the higher field as shown in Fig.~\ref{fig:phase}(c) or occupies only small area [Fig.~\ref{fig:phase}(a)]. This can be controlled by the cubic coupling $b$. For $\bm{H} \parallel [110]$, there is no high-$T$ phase in contrast to the case for $\bm{H} \parallel [001]$ in the experiments \cite{Onimaru2011-of,Kittaka2024-cn}. This point is consistent with our results. Other anomalies observed in the ultrasound experiments \cite{Ishii2011-mj} similar to those for $\bm{H} \parallel [001]$ are also interesting to examine in more detail in the future works. Since the anomalies are tiny and thus the comparison between the experiments and the present theory is not straightforward, while there is a phase boundary between the phase I$_{2q}$ and IV in Figs.~\ref{fig:phase2q4q_simulation}(c), \ref{fig:phase2q4q_simulation}(d), and \ref{fig:phase2q4q_general}(b), irrespectively to the detail of the parameters.

We also note that the phase diagram as a function of magnetic field from $\bm{H} \parallel [001]$ to $\bm{H} \parallel [110]$ is described in a function of angle $\Theta$, where  $\bm{H}=\tfrac{1}{\sqrt{2}}H(\sin\Theta,\sin\Theta,\sqrt{2}\cos\Theta)$. In Eq.~(\ref{eq:magH}), the effective field to the quadrupole moments is written as  $\tilde{\bm{h}}=gH^2(3\cos^2\Theta-1,0)$. Thus, the critical field of quadrupole orders for the angle $\Theta$, $H_{c}(\Theta)$ is expected to be
\begin{align}
    H_{c}(\Theta)=\frac{\sqrt{2}H_{c}(0)}{\sqrt{3\cos^2\Theta-1}}.
\end{align}
Such behavior can be one of strong evidences for quadrupole orders and indeed have been observed recently \cite{Okamoto2025-eo}.

\subsection{Comparison with quadrupole orders at the $\Gamma$ point and the X points}
In the previous studies, quadrupole orders in Pr-based 1-2-20 compounds have been theoretically analyzed in several types of ordering wavenumbers. In the early stage, simple N\'eel type quadrupole orders \cite{Hattori2014,Hattori2016} are analyzed and $T$--$H$ phase diagrams show similar order parameter switching as discussed in the analyses of single-$\bq$ sates shown in Fig.~\ref{fig:phase}. This is because the form of the Landau's free energy for both cases are almost the same when the multiple-$\bq$ configurations are neglected. 

For PrTi$_2$Al$_{20}$, the experiments clarify the order is ferroic \cite{Sakai2011,Taniguchi2019-uo,Kittaka2020} and its unusual behavior in the quadrupole switching under magnetic fields is analyzed by field-induced interactions unique to the quadrupole system \cite{Taniguchi2019-uo}. In this study such induced interactions are not taken into account. We do not consider they are important in PrIr$_2$Zn$_{20}$ since they require specific electronic properties such as existence of small pocket Fermi surfaces. 

For PrV$_2$Al$_{20}$, there are two transitions at zero magnetic field. The high-temperature transition is considered to be a quadrupole order, while ferroic octupole moments emerge below the lower transition temperature \cite{Patri2019-fq,Ye2024-vd}. 
The Landau analysis for a model including the octupole moment in addition to the quadrupole moments \cite{Lee2018} and Monte Carlo simulations with extended interactions have been done \cite{Freyer2018,Freyer2020-ns}. These theories explain the finite octupole moment as a consequence of symmetry breaking of ferroic octupole order. The lack of microscopic observation of both quadrupole and octupole moments (even their ordering wavenumbers) prevents theoretical investigation from further analyses. As an alternative approach, it has been pointed out that if the high-temperature transition is that into triple-$\bq$ quadrupole order at the X points suggested by the band structure calculation \cite{Iizuka2022-od}, the lower temperature transition is expected to be triple-$\bq$ octupole order at the same wavenumbers with induced ferroic octupole moments \cite{Ishitobi2021}. 

The physics of the L point quadrupole order is distinct from the ferroic and the X-point quadrupole moments in particular in the absence of cubic invariant consisting solely of the order parameters since the sum of three ordering wavenumbers lead to $\bk_1+\bk_2+\bk_3\ne {\bm 0}$. Although the N\'eel type order parameters do not possess such cubic invariant, they have trivially no multiple-$\bq$ orders. In this sense, the present results including double-$\bq$ quadrupole orders and the phase transitions from/to the single-$\bq$ orders are unique properties for the L point quadrupole systems. 

\subsection{Future perspectives}
We have shown that single- and double-$\bq$ quadrupole orders are realized in our study as a phenomenological model for PrIr$_2$Zn$_{20}$. The orthogonal double-$\bq$ orders are favored by the effective fourth-order terms arising from the renormalization due to the X-point fields, while the quadruple ones are not. Although we are interested in the orders at the L points in this paper, quadruple-$\bq$ orders at incommensurate ordering wave numbers along [111] directions has been studied in cubic magnets, showing hedgehog type topological magnetic structure \cite{Okumura2020-cu,Okumura2022-lo}. In principle, such exotic phases are also possible for the orbital model discussed in this paper when the exchange coupling favors incommensurate ordering wave numbers by introducing interaction frustration or anisotropic interaction ignored in this study \cite{Hattori2024-sk}. 
In this sense, it is interesting to explore possibilities of incommensurate orders as the candidates for the high-$T$ and intermediate field phase instead of the phase I proposed in this paper as studied in magnetic orders \cite{Bergman2007,Gao2017,Rosales2022-uh}.

For quantitative comparison between the theoretical and experimental results, it is also needed to incorporate the actual crystal electric field states in the real 
materials. They vary under magnetic fields and thus the parameters in the free energy can change. Recent ultrasonic experiments \cite{Ishii2025} detected rotational effects and its microscopic analysis is also interesting. For theoretical or computational aspects, it is also interesting to take into account the fluctuation effects by using e.g., classical Monte Carlo simulation employed in the previous studies on Pr-based materials \cite{Hattori2016,Freyer2018,Freyer2020-ns}. The thermal fluctuation may alter the mean-field phase diagram and to check their stability is important issue as one of our future problems \cite{sasa2026}. For example, our analysis of mode-mode coupling is restricted to those with the X-point modes via the cubic terms. This may cause overestimation of double-$\bq$ orders. In this sense, the numerical simulation can clarify the effects of these fluctuations. The influence of the non-Fermi liquid properties \cite{Sakai2011,onimaru2016,Han2022-jc,Inui2020-lu} on the symmetry breaking, which is completely ignored in this study, is also nontrivial and important subject.

\section{Summary}\label{sec:sum}
We have analyzed quadrupole orders with its modulation characterized by the wavenumber at the L points. Their Landau free energy is constructed under magnetic fields $\bm{H}$. We find that the phase diagrams show strong anisotropy. We have clarified that $x^2$$-$$y^2$ type quadrupole order is stabilized along $\bm{H} \parallel [110]$, which is consistent with that proposed in the neutron scattering experiment. We also propose that the unidentified high-temperature phase for $\bm{H}\parallel [001]$ in the experiments is a collinear $2z^2$$-$$x^2$$-$$y^2$ order and the orthogonal double-$\bq$ configurations are stable at low fields. These results offer further experimental examination of the order parameters in PrIr$_2$Zn$_{20}$ and related compounds. 

\section*{Acknowledgment}
The authors thank T. Onimaru, T. Ishitobi, and S. Kittaka for fruitful discussion. This work was supported by JSPS KAKENHI 
(Grant No.~JP23K20824, No.~JP23H04866, and No.~JP23H04869).

\appendix

\begin{figure*}[t]
\begin{center}
\includegraphics[width=\textwidth]{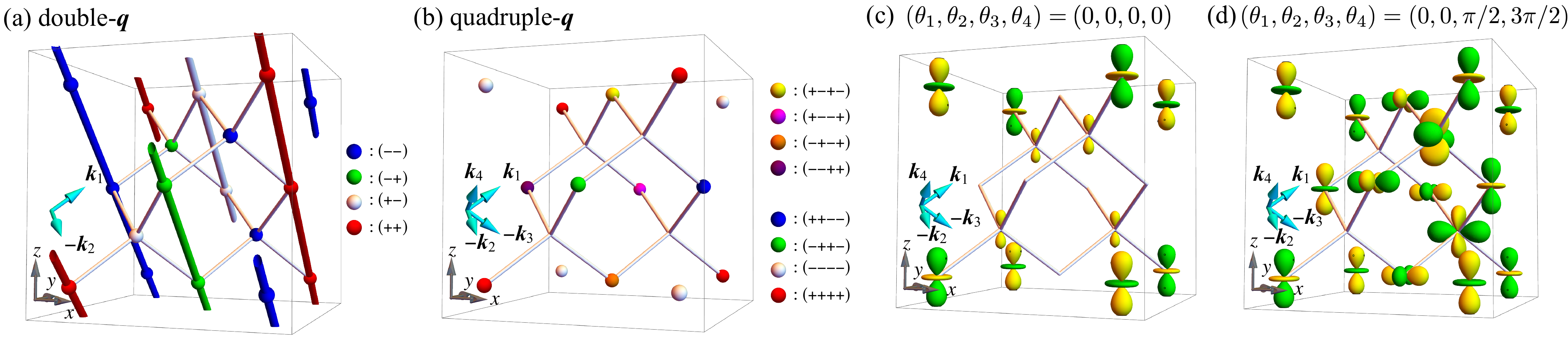}
\end{center}
\caption{(a) Double-$\bq$ configuration $\mathbb{E}_2$. Colored thick tubes connect the same $\mathbb{E}_2$'s. Thin bonds indicate the nearest-neighbor bonds of diamond structure. Arrows indicate the direction of $\bk_1$ and $-\bk_2$. (b) Quadruple-$\bq$ configuration $\mathbb{E}_4$, where only the $A$-sublattice sites are shown by spheres. (c) Symmetric quadruple-$\bq$ configuration for $(\theta_1,\theta_2,\theta_3,\theta_4)=(0,0,0,0)$ and (d) for $(\theta_1,\theta_2,\theta_3,\theta_4)=(0,0,\pi/2,3\pi/2)$. For (d), there are four kinds of ferroic chains. Two of them are similar to the double-$\bq$ configuration for $x=n$ ($n:$ integers) planes with the $2z^2$$-$$x^2$$-$$y^2$ orbital, while the other two chains for $x=n+\tfrac{1}{2}$ plane with the $x^2$$-$$y^2$ orbital run along $\bk_1+\bk_2$ direction. }
\label{fig:app_2q4q}
\end{figure*}

\section{Multiple-q configuration}\label{app:config}
In this Appendix, we show the real-space modulations for the double-$\bq$ and quadruple-$\bq$ orders since they are rather complicated. The real-space configurations for the double- and quadrupole-$\bq$ orders are given by Eq.~(\ref{eq:multiq_LC}). Here, we concentrate on the $A$-sublattice moments, since those for the $B$ sublattice are obtained by $\QQ_n\to \QQ_n(-1)^{p_n}$. Thus, one can rewrite Eq.~(\ref{eq:multiq_LC}) as 
\begin{align}
\bm{\phi}^A(\br)=\mathbb{E}_{\mathcal N}(\br)\cdot\begin{pmatrix}
	\bm{\phi}_1\\
	\bm{\phi}_2\\
	\cdots	
\end{pmatrix},\ 
\bm{\phi}_n=\frac{\Phi_n}{\sqrt{\mathcal N}}\begin{pmatrix}
	\cos\theta_n\\
	\sin\theta_n
\end{pmatrix},
\end{align}
where $n=1,2,\cdots, \mathcal{N}$ and $\mathbb{E}_{\mathcal N}$ is an ${\mathcal N}$ dimensional vector  
\begin{align}
	\mathbb{E}_{2}(\br)&=(e^{i\bk_1\cdot \br},e^{i\bk_2\cdot \br} ),\\
	\mathbb{E}_{4}(\br)&=(e^{i\bk_1\cdot \br},e^{i\bk_2\cdot \br},e^{i\bk_3\cdot \br},e^{i\bk_4\cdot \br}  ).
\end{align}
Here, we have assumed, without loss of generality, double-$\bq$ orders with $\bk_1$ and $\bk_2$. 
Since $\br$ is the position in the fcc lattice, there are four choices for $\mathbb{E}_{2}(\br) : (\pm 1,\pm 1)$, and eight choices for $\mathbb{E}_{4}(\br) : \pm (1,1,1,1),\ \pm (1,1,-1,-1),\ \pm (1,-1,1,-1)$, and $\pm (-1,1,1,-1)$. In Fig.~\ref{fig:app_2q4q}, $\mathbb{E}_{\mathcal{N}}(\br)$'s are illustrated and the colors distinguish the vectors $\mathbb E_{2,4}$. Note that only the $A$-sublattice positions in the diamond structure are shown in Fig.~\ref{fig:app_2q4q}(b). The unit cell for the double-$\bq$ orders contains four $A$-sublattice sites, each of which forms ferroic chain along the direction parallel to $\bk_3+\bk_4 \parallel (0,-1,1)$ as shown in Fig.~\ref{fig:app_2q4q}(a). The primitive translation vector along this direction is unchanged, while those perpendicular to it are modified to $\frac{1}{2}(2,1,1)$ and $\frac{1}{2}(2,-1,-1)$. 
For quadruple-$\bq$ orders, the primitive translation vectors are as twice large as those for the original fcc lattice: (1,1,0), (1,0,1), and (0,1,1). The unit cell contains eight sites for the quadruple-$\bq$ orders and sixteen when the $B$-sublattice sites are included.

Since several double-$\bq$ configurations are shown in Fig.~\ref{fig:config2q4q}, we here examine the real space configuration of the symmetric quadruple-$\bq$ orders discussed in Sec.~\ref{sec:symQ}, although they are not stabilized in the $T$--$H$ phase diagram in this paper.  It is evident that the configurations with ``collinear'' types such as 
$(\theta_1,\theta_2,\theta_3,\theta_4)=(0,0,0,0)$ and
$(\theta_1,\theta_2,\theta_3,\theta_4)=(0,0,0,\pi)$ are partially ordered states. For example, the quadrupole moments vanish at the sites where $\sum_ne^{i\bk_n\cdot \br}=0$ for the former as shown in Fig.~\ref{fig:app_2q4q}(c); $\boldsymbol{\phi}^A(\br)={\bf 0}$ for four fcc sites and $\boldsymbol{\phi}^B(\br)=-\tfrac{1}{2}\boldsymbol{\phi}^A(\bm{0})$. 
As an example of ``non-collinear'' case, Fig.~\ref{fig:app_2q4q}(d) illustrates the configuration for $(\theta_1,\theta_2,\theta_3,\theta_4)=(0,0,\pi/2,3\pi/2)$. This is also a partially ordered state, where $\boldsymbol{\phi}^B(\br)=\bm{0}$ at two of the $B$ sublattice sites within the cubic unit cell shown in Fig.~\ref{fig:app_2q4q}(d), while the other two have finite moments $|\boldsymbol{\phi}^B(\br)|=\sqrt{2}|\boldsymbol{\phi}^A(\bm{0})|$.

\section{Renormalization effects due to the fields at the X points}\label{app:reno}
In this Appendix, we consider the influence of the modes $\bQ_{\ell}^\alpha=(Q_{\ell,z}^\alpha,Q_{\ell,x}^\alpha)$ ($\ell=$ X, Y, Z) at the X points: 
$\bq_{\rm X}=(2\pi,0,0),\ \bq_{\rm Y}=(0,2\pi,0)$, and $\bq_{\rm Z}=(0,0,2\pi)$. They are more important than other modes at general wavenumbers since, for example, $\bq_{1}+\bq_{2}+\bq_{\rm X}\equiv \bf{0}$ leads to cubic mode coupling with the two different fields at the L points.   

  The free energy for the 
  X-points fields are sublattice diagonal and ${\mathcal F}_{\rm X}=\sum_{\alpha=A, B}{\mathcal F}_{\rm X}^\alpha$, where 
\begin{align}
{\mathcal F}^\alpha_{\rm X}=&\frac{1}{2}\sum_{\ell={\rm X,Y,Z}} 
\bQ^\alpha_{\ell}
\begin{bmatrix}
a_{\rm X}-6bu_0^\alpha& 6bv_0^\alpha\\
6b v_0^\alpha & a_{\rm X}+6bu_0^\alpha
\end{bmatrix}\bQ^\alpha_{\ell}\nonumber\\
&-6b \sum_{\ell={\rm X, Y, Z}}\sum_{i=1}^2\Big[Q_{\ell,z}^\alpha C_{\ell i}^\alpha - Q_{\ell,x}^\alpha S_{\ell i}^\alpha \Big]+\cdots.\label{Fx_only}
\end{align}
Here, the fourth-order terms and the cubic term for three X-point fields are neglected since the modes at the X points are not critical. $a_{\rm X}$ and $C$'s and $S$'s consisting of two L-point variables are given as 
\begin{align}
&a_{\rm X}\equiv T-4J_2+6J_4,\\
\begin{split}
	&C_{\ell 1}^\alpha=u_{m}^\alpha u_{n}^\alpha-v_{m}^\alpha v_{n}^\alpha,\quad 
 C_{\ell 2}^\alpha=u_{m'}^\alpha u_{n'}^\alpha-v_{m'}^\alpha v_{n'}^\alpha,\\ 
&S_{\ell 1}^\alpha=u_{m}^\alpha v_{n}^\alpha
+v_{m}^\alpha u_{n}^\alpha,\quad 
 S_{\ell 2}^\alpha=u_{m'}^\alpha v_{n'}^\alpha+v_{m'}^\alpha u_{n'}^\alpha,
 \end{split}
\end{align}
where $(\ell,m,n)=({\rm X},1,2),\ ({\rm Y},1,4),\ ({\rm Z},1,3)$, and $(\ell,m',n')=({\rm X},3,4),\ ({\rm Y},2,3),\ ({\rm Z},2,4)$.  
 By minimizing ${\mathcal F}_{\rm X}$ with respect to the X-point fields, we obtain their stationary values as 
\begin{align}
	&Q_{\ell,z}^\alpha=\frac{6 b}{a_{\rm X}}\sum_{i=1}^2 C^\alpha_{{\ell}i},
	\quad  Q^\alpha_{\ell,x}=-\frac{6 b}{a_{\rm X}}\sum_{i=1}^2 S^\alpha_{{\ell}i}.
\end{align}
 Substituting them into Eq.~(\ref{Fx_only}), we obtain up to the linear order in $\phi_{0}$: $\delta {\mathcal F}_{\rm {XL}}=\sum_{\alpha={\rm A,B}}\delta {\mathcal F}^\alpha_{\rm {XL}0}+\delta {\mathcal F}^\alpha_{\rm {XL}1}+\cdots$  as 
\begin{align}
	\delta {\mathcal F}^\alpha_{\rm {XL}0}=&-\frac{18b^2}{a_{\rm X}}\sum_{m>n}
(\phi^\alpha_{m}\phi^\alpha_{n})^2\nonumber\\	
&-\frac{36b^2}{a_{\rm X}}
(c_{12}^\alpha c_{34}^\alpha +c_{13}^\alpha c_{24}^\alpha+c_{14}^\alpha c_{23}^\alpha )\phi^\alpha_{1}\phi^\alpha_{2}\phi^\alpha_{3}\phi^\alpha_{4}, \label{eq:deltaF0}
\end{align}
and 
\begin{align}
	\delta {\mathcal F}_{\rm {XL}1}^\alpha=&-\frac{108 b^3}{a_{\rm X}^2}\Bigg[
\sum_{m>n}\cos(2\theta^\alpha_{mn}-\theta_0^\alpha)(\phi^\alpha_{m}\phi^\alpha_{n})^2\phi_0^\alpha
	\nonumber\\
	&+6\cos(\bar{\theta}^\alpha-\theta_0^\alpha)\phi^\alpha_1\phi^\alpha_2\phi^\alpha_3\phi^\alpha_4\phi_0^\alpha\Bigg],\label{eq:deltaF1}
\end{align}
where $\bm{\phi}_0^\alpha=(u_0^\alpha,v_0^\alpha)=\phi^\alpha_0(\cos\theta^\alpha_0,\sin\theta^\alpha_0)$ and $\theta_{mn}\equiv \theta_m+\theta_n$ and $\bar{\theta}^{\alpha}\equiv \theta_1^\alpha+\theta_2^\alpha+\theta_3^\alpha+\theta_4^\alpha$. 
Equation (\ref{eq:deltaF0}) renormalizes $c'$ and $c'''$ as 
\begin{align}
	c'\to c'-\frac{72b^2}{a_{\rm X}}, \quad	c'''\to c'''-\frac{144b^2}{a_{\rm X}}.\label{eq:app_reno}
\end{align}
The terms appearing in $\delta {\mathcal F}_{\rm {XL}1}^\alpha$ are emergent fifth-order coupling constants between the mode at the $\Gamma$ and the L points. Since we have kept terms up to the fourth orders in the original free energy in Eq.~(\ref{eq:FLtot}), cases for large $108b^3/a_{\rm X}$ are beyond the applicable range of the approximation.  We also note that the last terms in Eqs.~(\ref{eq:deltaF0}) and (\ref{eq:deltaF1}) vanish  when the sublattice ($\alpha$) summations are carried outwith assuming $\bm{\phi}_n^B=(-1)^{p_n}\bm{\phi}_n^A$. See the discussions in Secs.~\ref{sec:IIIA} and \ref{sec:symQ}.

\section{Free energy for the orthogonal double-q configuration} \label{sec:appC}
We show the free energy for the orthogonal double-$\bq$ orders with $\bm{\phi}_2\perp \bm{\phi}_1$: $\bm{\phi}_1=\phi_1(\cos\theta,\sin\theta), \bm{\phi}_2=\phi_2(\cos(\theta+\tfrac{\pi}{2}),\sin(\theta+\tfrac{\pi}{2}))=\phi_2(-\sin\theta,\cos\theta)$, with the $\Gamma$-point mode $\bm{\phi}_0=\phi_0(\cos\theta_0,\sin\theta_0)$. Here, we assume $\bm{\phi}_1=\bm{\phi}_{1}^A=\bm{\phi}_{1}^B$ and $\bm{\phi}_2=\bm{\phi}_{2}^A=-\bm{\phi}_{2}^B$ as discussed in the main text. The free energy is given as 
\begin{align}
{\mathcal F}_{\rm or}=&\sum_{n=1,2}\Big[(a_{\rm L}+J_{\rm L} +\lambda_c^{(1)}\phi_0^2)\phi_n^2+\frac{c}{2}\phi_n^4\Big]+\frac{c'}{2}\phi_1^2\phi_2^2\nonumber\\
	&-6\lambda_b\phi_0(\phi_1^2-\phi_2^2)\cos(2\theta+\theta_0)\nonumber\\
	&+2\lambda_c^{(2)} \phi_0^2 \Big[
	\phi_1^2\cos^2(\theta-\theta_0)+	\phi_2^2\sin^2(\theta-\theta_0)
	\Big] \nonumber\\
	&+\frac{216 b^3}{a_{\rm X}^2}
\phi_{0}\phi^2_{1}\phi^2_{2}\cos(4\theta-\theta_0)+{\mathcal F}_\Gamma, 
\label{eq:appendixB1}
\end{align}
where $c'=2c-72b^2/a_{\rm X},\ \lambda_b=b,\ \lambda_c^{(1)}=\lambda_c^{(2)}=c$, and ${\mathcal F}_\Gamma$ is given by Eq.~(\ref{eq:F2Geff}).
Let us briefly discuss possible stationary angle $\theta$ under magnetic fields. For simplicity, we assume $\theta_0=0$, since $\tilde{h}$ induces $u_0$ under both $\bm{H}\parallel [001]$ and $[110]$. We treat $\phi_{0,1,2}$ as given parameters and the stationary condition of ${\mathcal F}_{\rm or}$ with respect to $\cos(2\theta)$ leads to
\begin{align}
\cos(2\theta)=\frac{6\lambda_b-\lambda_c^{(2)}\phi_0}{216b^3/a_{\rm X}^3 }\left(\frac{\phi_1^2-\phi_2^2}{\phi_1^2\phi_2^2}\right).
\end{align}
This means that $\phi_1/\phi_2\to 1$ for $\theta\to \pi/4$ unless $\phi_0$ is tuned to $6\lambda_b/\lambda_c^{(2)}$, which is the phase III$_{2q}$ seen in the main text. 
The free energy for the phase III$_{2q}$ is 

\begin{align}
	{\mathcal F}_{{\rm III}_{2q}}=&2\big[a_{\rm L}+J_{\rm L} +(\lambda_c^{(1)}+\lambda_c^{(2)})\phi_0^2\big]\phi_1^2
\nonumber\\
	&+\left( c +\frac{c'}{2} -\frac{216 b^3}{a_{\rm X}^2}
\phi_{0} \cos\theta_0\right)\phi_1^4 
	+{\mathcal F}_\Gamma.
\end{align}
Although we do not discuss this in detail, 
we can also see that the deviation $\delta\phi=\phi_1-\phi_2\ne 0$ from the phase III$_{2q}$ accompanies changes in the angle as $\theta=\pi/4+\epsilon$. The quadratic form of $\delta\phi$ and $\epsilon$ describes the stability of the phase III$_{2q}$. When one of the eigenvalues vanishes, a second-order transition takes place toward the phase II$_{2q}$.
	

For $\theta=0$, i.e., the phase I$_{2q}$, the free energy is given by 
\begin{align}
	{\mathcal F}_{{\rm{I}}_{2q}}=&\sum_{n=1,2}\Big[(a_{\rm L}+J_{\rm L} +\lambda_c^{(1)}\phi_0^2)\phi_n^2+\frac{c}{2}\phi_n^4\Big]+\frac{c'}{2}\phi_1^2\phi_2^2\nonumber\\
	&-6\lambda_b\phi_0(\phi_1^2-\phi_2^2)\cos\theta_0\nonumber\\
	&+2\lambda_c^{(2)} \phi_0^2 \Big[
	\phi_1^2\cos^2\theta_0+	\phi_2^2\sin^2\theta_0
	\Big] \nonumber\\
	&+\frac{216 b^3}{a_{\rm X}^2}
\phi_{0}\phi^2_{1}\phi^2_{2}\cos\theta_0+{\mathcal F}_\Gamma.
\end{align}
Rotating $\theta$ leads to the transition to the phase II$_{2q}$. This accompanies also rotation in $\theta_0$ accordingly.

%

\end{document}